\documentclass[lettersize,journal]{IEEEtran}

\usepackage{amsmath,amsfonts}
\usepackage{algorithmic}
\usepackage{algorithm}
\usepackage{array}
\usepackage[caption=false]{subfig}
\usepackage{textcomp}
\usepackage{stfloats}
\usepackage{url}
\usepackage{verbatim}
\usepackage{graphicx}
\usepackage{cite}
\usepackage{xcolor}
\usepackage[acronym]{glossaries} 
\usepackage{booktabs} 
\usepackage{multirow} 
\usepackage[hidelinks]{hyperref} 
\usepackage{amssymb}  
\usepackage{microtype}
\usepackage{makecell}
\usepackage[inline]{enumitem}
\usepackage{scalefnt}
\usepackage{flushend}

\makeglossaries
\newacronym{3gpp}{3GPP}{3rd Generation Partnership Project}
\newacronym{aod}{AoD}{angle-of-departure}
\newacronym{ap}{AP}{antenna point}
\newacronym{awgn}{AWGN}{additive white Gaussian noise}
\newacronym{bie}{BIE}{best integer equivariant}
\newacronym{bpsk}{BPSK}{binary phase-shift keying}
\newacronym{cdf}{CDF}{cumulative distribution function}
\newacronym{cf}{CF}{cell-free}
\newacronym{cnn}{CNN}{convolutional neural network}
\newacronym{conv2d}{Conv2D}{2D-Convolutional}
\newacronym{cpo}{CPO}{carrier phase offset}
\newacronym{cpprs}{CP-PRS}{carrier phase positioning reference signal}
\newacronym{cpp}{CPP}{carrier phase positioning}
\newacronym{csi}{CSI}{channel state information}
\newacronym{dmimo}{D-MIMO}{distributed multiple-input multiple-output}
\newacronym{dl}{DL}{deep learning}
\newacronym{dlmf}{DLMF}{distributed location management function}
\newacronym{dmrs}{DMRS}{demodulation reference signal}
\newacronym{dnn}{DNN}{deep neural network}
\newacronym{ecdf}{ECDF}{empirical cumulative distribution function}
\newacronym{egs}{EGS}{exhaustive grid search}
\newacronym{ekf}{EKF}{extended Kalman ﬁlter}
\newacronym{efim}{EFIM}{equivalent Fisher information matrix}
\newacronym{fcnn}{FCNN}{fully connected neural network}
\newacronym{ffn}{FFN}{feed-forward network}
\newacronym{fim}{FIM}{Fisher information matrix}
\newacronym{flop}{FLOP}{floating-point operation}
\newacronym{fc}{FC}{fully connected}
\newacronym{gd}{GD}{gradient descent}
\newacronym{gelu}{GELU}{Gaussian Error Linear Unit}
\newacronym{gap}{GAP}{global average pooling}
\newacronym[
    longplural={Gaussian processes},
    shortplural={GPs}
]{gp}{GP}{Gaussian process}
\newacronym{gpr}{GPR}{Gaussian Process Regression}
\newacronym{gr}{GR}{ground reflection}
\newacronym{gnss}{GNSS}{global navigation satellite system}
\newacronym{hi}{HI}{hyperbola intersection}
\newacronym{hrllc}{HRLLC}{Hyper Reliable and Low Latency Communication}
\newacronym{ln}{LN}{Layer Norm}
\newacronym{lrelu}{Leaky ReLU}{Leaky Rectified Linear Unit}
\newacronym{lmf}{LMF}{location management function}
\newacronym{los}{LoS}{line-of-sight}
\newacronym{lr}{LR}{learning rate}
\newacronym[
    longplural={long short-term memories},
    shortplural={LSTMs}
]{lstm}{LSTM}{long short-term memory}\newacronym{mae}{MAE}{mean absolute error}
\newacronym{mcs}{MCS}{Monte Carlo simulation}
\newacronym{mimo}{MIMO}{multiple-input multiple-output}
\newacronym{mli}{ML}{maximum-likelihood}
\newacronym{mle}{MLE}{maximum-likelihood estimation}
\newacronym{mlp}{MLP}{multi-layer perceptron}
\newacronym{mse}{MSE}{mean-squared error}
\newacronym{mha}{MHA}{multi-head attention}
\newacronym{gnb}{gNB}{next-generation NodeB}
\newacronym{nlos}{NLoS}{non-line-of-sight}
\newacronym{nn}{NN}{neural network}
\newacronym{nr}{NR}{New Radio}
\newacronym{ofdm}{OFDM}{orthogonal frequency-division multiplexing}
\newacronym{popt}{POPT}{Phase-Only Positioning Transformer}
\newacronym{peb}{PEB}{position error bound}
\newacronym{prs}{PRS}{positioning reference signal}
\newacronym{psd}{PSD}{power spectral density}
\newacronym{rcs}{RCS}{radar cross section}
\newacronym{relu}{ReLU}{Rectified Linear Unit}
\newacronym{rmse}{RMSE}{root mean squared error}
\newacronym{slam}{SLAM}{simultaneous localization and mapping}
\newacronym{snr}{SNR}{signal-to-noise ratio}
\newacronym{srs}{SRS}{sounding reference signal}
\newacronym{tdoa}{TDoA}{time-difference-of-arrival}
\newacronym{te}{TE}{transverse electric}
\newacronym{toa}{ToA}{time-of-arrival}
\newacronym{tof}{ToF}{time-of-flight}
\newacronym{ue}{UE}{user equipment}
\newacronym{uwb}{UWB}{ultra-wideband}
\newacronym{vit}{ViT}{Vision Transformer}
\newacronym{rhs}{RHS}{right-hand side}
\newacronym{nlpd}{NLPD}{negative log predictive density}
\newacronym{ls}{LS}{least squares}
\newcommand{\N}[0]{\ensuremath{\mathcal{N}}}
\newcommand{\GP}[0]{\ensuremath{\mathcal{GP}}}

\newcommand{\acos}{ \textrm{acos}  }
\newcommand{\argmin}{ \textrm{arg\,min\,}  }

\newcommand{\diag}{ \textrm{diag} }
\newcommand{\trace}{ \textrm{Tr} }

\newcommand{\appropto}[0]{%
    \ensuremath{%
        \mathrel{%
            \vcenter{%
                \offinterlineskip\halign{%
                    \hfil$##$\cr%
                    \propto\cr\noalign{\kern2pt}%
                    \sim\cr\noalign{\kern-2pt}%
                }%
            }%
        }%
    }%
}

\DeclareMathOperator{\V}{V}
\DeclareMathOperator{\E}{E}

\renewcommand{\vec}[1]{\ensuremath{{\mathbf{#1}}}}
\newcommand{\vecsymbol}[1]{\ensuremath{\boldsymbol{#1}}}

\newcommand{\vb}[0]{\vec{b}}

\newcommand{\vf}[0]{\vec{f}}

\newcommand{\vh}[0]{\vec{h}}

\newcommand{\vk}[0]{\vec{k}}

\newcommand{\vp}[0]{\vec{p}}

\newcommand{\vr}[0]{\vec{r}}
\newcommand{\vs}[0]{\vec{s}}
\newcommand{\vt}[0]{\vec{t}}

\newcommand{\vx}[0]{\vec{x}}
\newcommand{\vy}[0]{\vec{y}}

\newcommand{\vE}[0]{\vec{E}}

\newcommand{\vH}[0]{\vec{H}}
\newcommand{\vI}[0]{\vec{I}}
\newcommand{\vJ}[0]{\vec{J}}
\newcommand{\vK}[0]{\vec{K}}

\newcommand{\vP}[0]{\vec{P}}

\newcommand{\vT}[0]{\vec{T}}

\newcommand{\vW}[0]{\vec{W}}

\newcommand{\valpha}[0]{\vecsymbol{\alpha}}

\newcommand{\vnu}[0]{\vecsymbol{\nu}}
\newcommand{\veta}[0]{\vecsymbol{\eta}}
\newcommand{\vupsilon}[0]{\vecsymbol{\upsilon}}
\newcommand{\vtheta}[0]{\vecsymbol{\theta}}

\newcommand{\vpsi}[0]{\vecsymbol{\psi}}
\newcommand{\vphi}[0]{\vecsymbol{\phi}}

\newcommand{\vLambda}[0]{\vecsymbol{\Lambda}}

\newcommand{\cI}[0]{\mathcal{I}}

\newcommand{\cL}[0]{\mathcal{L}}
\newcommand{\cX}[0]{\mathcal{X}}
\newcommand{\cM}[0]{\mathcal{M}}

\DeclareMathOperator{\atan2}{atan2}

\newcommand{\vdeltam}{\boldsymbol{\delta}_m}
\newcommand{\vdeltamtilde}{\tilde{\boldsymbol{\delta}}_m}

\allowdisplaybreaks

\begin{document}
\bstctlcite{IEEEexample:BSTcontrol}   
\scalefont{0.97}

\title{POPT: Physics-Informed Deep Learning for\\Phase-Only Positioning in Distributed MIMO}

\author{Fatih Ayten,~\IEEEmembership{Graduate Student Member,~IEEE}, Ossi Kaltiokallio,~\IEEEmembership{Member,~IEEE}, Jukka Talvitie,~\IEEEmembership{Member,~IEEE}, Akshay Jain,~\IEEEmembership{Senior Member,~IEEE}, Mehmet C. Ilter,~\IEEEmembership{Senior Member,~IEEE}, \\  Musa Furkan Keskin,~\IEEEmembership{Member,~IEEE}, Elena Simona Lohan,~\IEEEmembership{Senior Member,~IEEE}, \\Henk Wymeersch,~\IEEEmembership{Fellow,~IEEE}, and Mikko Valkama,~\IEEEmembership{Fellow,~IEEE} 
\thanks{F. Ayten, O. Kaltiokallio, M. C. Ilter, J. Talvitie, E. S. Lohan, and M. Valkama are with 
Tampere University, 
Finland. 
}
\thanks{A. Jain is with Nokia Bell Labs, Espoo, Finland. 
} 
\thanks{M. F. Keskin and H. Wymeersch are with 
Chalmers University of Technology, 
Sweden. 
}
\vspace{-4mm}
}

\maketitle

\begin{abstract}
\textls[-1]{This article develops a physics-informed deep learning framework for single-snapshot 3D phase-only positioning in narrowband phase-coherent \gls{dmimo} networks under two-ray propagation. Unlike prior phase-coherent \gls{dmimo} localization methods that largely assume \gls{los}-only channels, we consider a \gls{los} path and a specular ground reflection. Since the narrowband single-snapshot observation does not provide sufficient delay resolution to separate these components, their coherent superposition induces structured perturbations in the carrier phase measurements. To address this, we first develop a \gls{gp}-based model that learns the quasi-periodic phase distortion caused by ground reflection from carrier phase measurements collected at a small number of training locations, and then uses it to generate high-quality synthetic samples. Building on these samples, we propose \gls{popt}, an encoder-only transformer for phase-only localization that captures inter-\gls{ap} phase relationships without solving the highly non-convex \gls{mli} problem underlying model-based estimators. We also derive the fundamental \gls{peb} and {develop} the \gls{mle} for the considered multipath phase-only \gls{dmimo} positioning problem. Numerical results show that, with only 50 training locations for the \gls{gp} model, the proposed method achieves near-\gls{peb} accuracy. {The results further highlight that the analytical \gls{mle} is highly sensitive to relative permittivity mismatch, whereas the proposed method mitigates this dependence by learning the phase perturbation directly from measurement data.} Furthermore, compared to the \gls{mle}, the proposed approach reduces \gls{flop} complexity and inference time by about 1.7 and 3.6 \emph{orders of magnitude}, respectively. 
}
\end{abstract}

\glsresetall    
\vspace{-2mm}

\vspace{-4mm}
\section{Introduction}
\vspace{-1mm}
\textls[-5]{While 5G-Advanced can provide positioning accuracy of around 10\,cm under favorable conditions, this accuracy must be significantly improved for 6G use cases such as cooperative mobile robots, real-time digital twins, and remote construction~\cite{3gpp_tr_22_870}. Conventional positioning methods based on time- and angle-based measurements, such as \gls{tdoa} and \gls{aod}, are effective, but their performance is bounded by physical system resources. Specifically, bandwidth limits the resolution of time-based measurements, while antenna aperture and array resolution limit angle-based estimation, thereby constraining the final positioning accuracy~\cite{11270931}. \Gls{cpp}, which is well established in the \gls{gnss} context~\cite{ding2021carrier}, has emerged as a promising candidate for centimeter-level terrestrial positioning~\cite{10475845,9566601}. Its adoption in \gls{3gpp} standardization has also gained traction, particularly in 5G-Advanced, where carrier phase measurements are being explored as an emerging means to enhance positioning capabilities~\cite{10536135}. Nevertheless, \gls{cpp} is subject to so-called integer ambiguity challenge, since the mapping between the transmitter--receiver distance and the observable carrier phase remains invariant to integer multiples of the wavelength~\cite{10232971}.}

\begin{figure}[t!]
    \centering
  \includegraphics[width=0.48\textwidth]{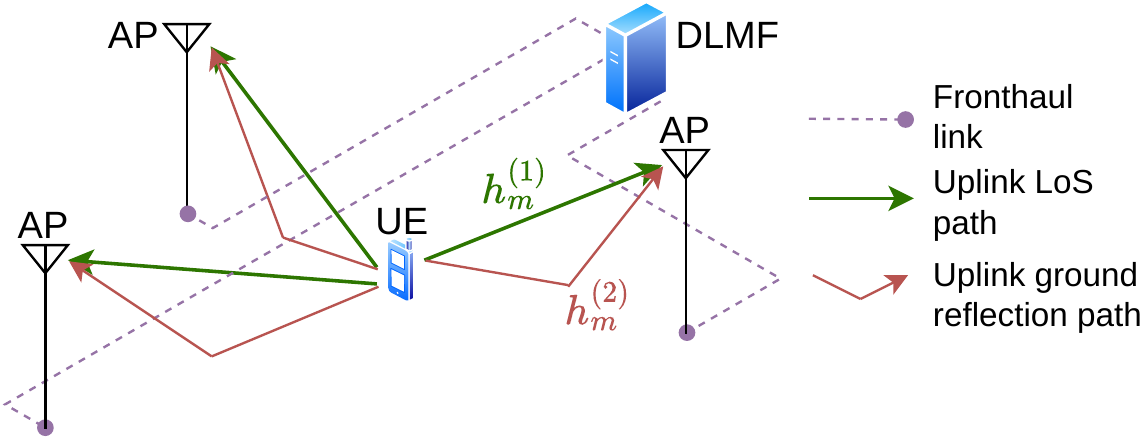}
  \vspace{-1mm}
  \caption{Illustration of the considered uplink phase-only positioning scenario, where distributed \glsentrylongpl{ap} (\glsentryshortpl{ap}) forward measurements to a coordinating \glsentrylong{dlmf} (\glsentryshort{dlmf}). Only carrier phase measurements are taken into account for \glsentrylong{ue} (\glsentryshort{ue}) positioning.}
  \vspace{3mm}
  \label{fig:system_figure}
\end{figure}
\glsreset{ue}
\glsreset{ap}


\vspace{-4mm}
\subsection{Related Works and CPP Prior Art}
\textls[-7]{In the context of 5G and 5G-Advanced \gls{nr} mobile networks, \gls{cpp} through integer ambiguity resolution has been recently studied in~\cite{9601204, 10012896, 10437192, 9566601, 10232971, s21206731, 10139995, 9952844, 9698115, 9977723, 10437902, 10901681, shourezari2026multibandcarrierphasepositioning}. The first complete \gls{cpp} framework for cellular networks is presented in~\cite{9566601}, focusing on clock synchronization and integer ambiguity resolution. The proposed framework fuses \gls{tdoa}-based \gls{ue} position estimates with the temporal changes of carrier phase measurements. The approach in~\cite{9601204} refines \gls{toa} ranging by estimating the carrier phase of the first-arriving path. A carrier phase-based ranging scheme is proposed in~\cite{s21206731}, employing an \gls{ekf}-based algorithm for integer ambiguity resolution and \gls{nlos} error mitigation. A single-snapshot positioning method integrating \gls{cpp} and \gls{tdoa} is developed in~\cite{10901681}, where \gls{tdoa} measurements are used for preliminary positioning of the \gls{ue}. Notably,~\cite{10437902} introduces a novel bound that captures the impact of integer ambiguity on \gls{ue} positioning based on time and carrier phase measurements. More recently,~\cite{shourezari2026multibandcarrierphasepositioning} extends this line of work to multi-band \gls{cpp}, deriving corresponding performance bounds and efficient estimators. A wideband \gls{cpprs} waveform is employed in~\cite{9977723} to achieve high-precision positioning, assuming perfect resolution of the integer ambiguity problem, and the sensitivity of the positioning performance to integer ambiguity errors is analyzed. In~\cite{10012896}, multi-frequency carrier phase measurements are linearly combined to form a virtual frequency, which mitigates the impact of \gls{ap} position errors and facilitates integer ambiguity resolution. Triple-frequency ambiguity resolution is proposed in~\cite{10437192}, where ambiguities are resolved by combining observations across three frequencies. Similarly,~\cite{9952844} employs \gls{ofdm} waveforms, where multiple subcarriers are leveraged to resolve the integer ambiguity. A shared limitation of~\cite{9566601, 9601204, s21206731, 10901681} is that carrier phase measurements are treated as an add-on to \gls{toa} or \gls{tdoa} observations, which often demands high bandwidth~\cite{1259415, 1458289} and, consequently, higher system resource utilization. By contrast, the approaches in~\cite{9977723, 10012896, 10437192, 9952844, shourezari2026multibandcarrierphasepositioning} do not use \gls{toa}/\gls{tdoa} as the main positioning observable; instead, they rely on multiple subcarriers to sustain high positioning accuracy, thereby requiring additional frequency-domain resources compared with single-subcarrier phase-only positioning.}

An alternative strategy for integer ambiguity resolution is to combine observations collected over multiple snapshots. For example, an \gls{ekf}-based algorithm in~\cite{10232971} operates over multiple snapshots to mitigate \gls{ue} clock drift and resolve integer ambiguities; continuous transmission of \glspl{prs} in~\cite{10139995} enables accurate carrier phase tracking and helps prevent integer ambiguity occurrences; and a factor graph framework in~\cite{9698115} leverages time-differential carrier phase measurements for positioning. However, approaches that rely on multiple snapshots~\cite{10232971, 10139995, 9698115} typically operate in a history-dependent tracking setting, where the estimation process exploits temporal continuity and previously available state information, rather than addressing single-snapshot positioning.

A promising alternative approach for high-precision localization is direct positioning~\cite{11226252, 10851402, s19204582, Vukmirovic2018PositionEstimation, 7849233, s21103401}, which estimates the source location in a single snapshot directly from the raw measurements (e.g., via \gls{mle}), rather than first estimating intermediate parameters such as \gls{toa}. However, coherent processing typically yields a spiky, highly non-convex \gls{mle} cost function with many local minima near the true \gls{ue} position, and coherent direct localization may require an extremely fine search grid, increasing computational complexity. To alleviate this issue, multi-stage and coarse-to-fine search strategies have been proposed~\cite{11226252, 10851402, Vukmirovic2018PositionEstimation} to reduce the sensitivity to local minima while retaining most of the coherent gain. Overall, \gls{cpp} in spatially distributed coherent networks can provide substantial accuracy improvements over non-coherent positioning approaches, but at the cost of a more challenging estimation problem.


\textls[-4]{In \emph{\gls{dmimo}} architectures, illustrated conceptually in Fig.~\ref{fig:system_figure}, multiple geographically distributed \glspl{ap} observe the \gls{ue} from different spatial viewpoints, which can improve coverage uniformity, spatial diversity, and positioning geometry compared with conventional collocated deployments~\cite{7827017}. Under phase-coherent operation, the distributed \glspl{ap} effectively form a large sparse aperture, and the resulting wavefront-curvature information can provide an additional source of spatial discrimination~\cite{10851402}. These properties make phase-coherent \gls{dmimo} particularly attractive for phase-only positioning, where localization relies on carrier phase observations rather than delay resolution and can therefore operate with minimal bandwidth requirements~\cite{11275391, 11274977, fatih_icassp, dey2025polo}. In our prior works~\cite{11275391, 11274977, fatih_icassp}, we proposed \gls{dl}-based phase-only positioning approaches. More recently, a low-complexity localization framework was proposed in~\cite{dey2025polo} that exploits geometric constraints induced by selected \gls{ap} pairs to substantially reduce the complexity of the \gls{mle} search. Closely related,~\cite{yu_ge_arxiv} studies uplink single-snapshot frugal \gls{slam} in phase-coherent \gls{dmimo} systems with narrowband single-subcarrier measurements and single-antenna \glspl{ap}, jointly detecting and localizing the \gls{ue}, reflective surfaces, and scatterers through coherent imaging and iterative interference cancellation. However, the prior machine learning-based approaches in~\cite{11275391, 11274977, fatih_icassp} typically require a large number of labeled training samples, which can be impractical to acquire over large coverage areas. On the other hand, the coverage of~\cite{dey2025polo} depends strongly on the \gls{ap} deployment and is therefore not guaranteed. Moreover, the phase-only positioning approaches in~\cite{11275391, 11274977, fatih_icassp, dey2025polo} are restricted to \gls{los}-only propagation and 2D scenarios. Such a \gls{los}-only assumption is particularly restrictive in the narrowband phase-only regime, where multipath components are not resolvable in the delay domain and are therefore coherently superimposed with the \gls{los} component. Consequently, narrowband phase-only estimators designed under a \gls{los}-only model can suffer substantial performance degradation in the presence of specular multipath. Although~\cite{yu_ge_arxiv} considers a more general \gls{slam} setting with reflective surfaces and scatterers, it focuses on grid-based image formation and sequential target extraction rather than direct phase-only \gls{ue} positioning, and assumes known \gls{ue} height.}

\vspace{-3mm}
\subsection{Contributions and Novelty}
This article addresses the above-identified research gaps, and develops a physics-informed deep learning framework for single-snapshot 3D phase-only positioning in uplink narrowband phase-coherent \gls{dmimo} systems under two-ray propagation. The two-ray model is considered as an important first step beyond the commonly used \gls{los}-only assumption. In terrestrial deployments, the ground surface is almost always present and can create a strong specular reflection. For narrowband carrier phase positioning, the \gls{los} and ground-reflected components cannot be resolved in the delay domain; instead, they are coherently superimposed. Therefore, even this single dominant reflected path can severely challenge phase-only positioning methods designed under a \gls{los}-only model. Furthermore, the two-ray model remains structured to enable analytical performance bounds. Therefore, it provides a controlled but meaningful bridge between \gls{los}-only phase-only positioning and more general multipath propagation.
To the best of our knowledge, this is the first study of this setting, where prior works either assume \gls{los}-only channels~\cite{11226252, dey2025polo, 11275391, 11274977, fatih_icassp}, rely on wideband or multi-frequency signals~\cite{10851402, s19204582, shourezari2026multibandcarrierphasepositioning, s21103401}, or require multi-antenna \glspl{ap}~\cite{10851402, 11226252, Vukmirovic2018PositionEstimation, 7849233}. Specifically, we leverage \glspl{gp} to model the quasi-periodic phase perturbations induced by ground reflections in a two-ray propagation scenario, without requiring prior knowledge of the ground material or polarization type, both of which are required by methods relying on analytical models. Since the specular ground reflection is governed by the \gls{ap}-\gls{ue} geometry rather than by the \gls{ap} identity, we learn a single shared \gls{gp} model across all \glspl{ap} from carrier phase measurements obtained from a limited number of labeled samples. As each labeled \gls{ue} location provides observations from multiple \glspl{ap}, the effective number of training observations scales with the number of \glspl{ap}, thereby substantially reducing the number of labeled samples required. The proposed \gls{gp} framework enables the generation of high-quality synthetic training samples from this limited labeled dataset, which in turn allows the subsequent learning-based positioning stage to be trained without requiring dense labeled samples over the coverage area, significantly reducing the data requirements compared to prior works~\cite{11275391, 11274977, fatih_icassp}. Building on this, we propose and design \gls{popt}, an encoder-only transformer architecture trained with the synthetic dataset produced by the \gls{gp} model for accurate 3D \gls{ue} positioning. 

The specific contributions can be stated as follows:
\begin{itemize}[
    topsep=0pt,
    itemsep=0pt,
    parsep=0pt,
    partopsep=0pt,
    labelindent=\parindent,
    leftmargin=*
]
\item \textbf{Analytical model, performance bounds, and \gls{mle} development:} 
We introduce the underlying analytical model for the considered positioning problem and use it to establish the theoretical foundation of the article. Based on this model, we derive the phase-only \gls{peb}. We further develop a phase-only \gls{mle} based on \gls{egs} as a model-based solution. The developed \gls{mle} serves as a theoretically grounded but computationally complex baseline for evaluating the proposed learning-based approach.

\item \textbf{Shared \gls{gp}-based ground reflection modeling and synthetic data generation:} 
\textls[-3]{We propose a novel \gls{gp} framework that captures the quasi-periodic phase perturbations induced by ground reflection. A key feature of the proposed model is that the perturbations associated with all \glspl{ap} are explained by a single shared \gls{gp} model, rather than by independent \gls{ap}-specific models. This shared structure exploits the common physical origin of the perturbations and allows the model to be trained from carrier phase measurements obtained from only a limited number of labeled samples. The trained \gls{gp} is then used to generate high-quality synthetic phase-only samples, thereby reducing the need for dense labeled training data.}

\item \textbf{Transformer-based phase-only positioning:}
Building on the synthetic samples generated by the shared \gls{gp} model, we develop \gls{popt}, the first encoder-only transformer architecture for single-snapshot 3D phase-only positioning. The proposed network uses self-attention to capture inter-\gls{ap} phase relationships while avoiding the high computational complexity of \gls{mle}-based methods. This creates a direct link between the data-efficient \gls{gp} model and the learning-based positioning stage, where a limited number of labeled samples is used to learn the perturbation model, and the resulting synthetic data enables accurate, low-complexity inference with \gls{popt}.
\end{itemize}

For clarity, we note that data-driven learning of environment-dependent radio features has emerged as a promising direction in the literature. In~\cite{app10155216}, \gls{gp} regression is employed to model the angle-dependent amplitude and phase behavior of specular multipath components in \gls{uwb} and mm-wave settings. In contrast to~\cite{app10155216}, the present work considers narrowband \gls{dmimo} carrier phase positioning, where the \gls{los} and reflected components are not resolvable in the delay domain. More recently, the authors in~\cite{chatelier2026} proposed a model-based neural network that learns the location-to-channel mapping and serves as a generative neural channel model.

The remainder of this article is organized as follows. In Section~\ref{sec:system_model}, we present the system model and formulate the phase-only positioning problem. We then derive and introduce the baseline \gls{mle} and provide fundamental performance bounds. Section~\ref{sec:gaussian_process} introduces the proposed \gls{gp}-based carrier phase model, capturing the underlying physical characteristics of the propagation environment. In Section~\ref{sec:positioning}, we develop the proposed positioning approach, \gls{popt}, and describe the benchmark methods. The computational complexities of the considered approaches are described in Section~\ref{sec:complexity_analysis}. Section~\ref{sec:num_results} presents the numerical results. Finally, Section~\ref{sec:conclusion} contains our concluding remarks, {while selected derivations are provided in the Appendix}. 

\textit{Notations:} \textls[-3]{Scalars, vectors, and matrices are denoted by italic lowercase, bold lowercase, and bold uppercase letters, respectively. The sets of real and complex numbers are denoted by $\mathbb{R}$ and $\mathbb{C}$. The $i$th element of a vector $\vx$ is denoted by $[\vx]_i$, and $\diag(\vx)$ denotes the diagonal matrix with the entries of $\vx$ on its main diagonal. The operators $(\cdot)^\top$ and $(\cdot)^{-1}$ denote transpose and matrix inverse, respectively. The magnitude and phase of a complex scalar are denoted by $\lvert \cdot \rvert$ and $\angle(\cdot)$, respectively, and $\jmath \triangleq \sqrt{-1}$ denotes the imaginary unit. Furthermore, $\exp(\cdot)$, $\trace(\cdot)$, and $\acos(\cdot)$ denote the exponential, trace, and inverse cosine operators, respectively. The Euclidean norm and the $\ell_1$-norm are denoted by $\lVert \cdot \rVert_2$ and $\lVert \cdot \rVert_1$, respectively, while $\odot$ denotes element-wise multiplication. The distributions $\mathcal{N}(\boldsymbol{\mu},\mathbf{\Sigma})$ and 
$\mathcal{CN}(\boldsymbol{\mu},\mathbf{\Sigma})$ denote real and circularly 
symmetric complex Gaussian distributions, respectively, while 
$\GP(0,k(\cdot,\cdot))$ denotes a zero-mean \gls{gp} prior with covariance 
function $k(\cdot,\cdot)$. The operators \(\Re\{\cdot\}\) and \(\Im\{\cdot\}\) denote the real and imaginary parts, respectively.
The expectation and variance operators are denoted by \(\E[\cdot]\) and \(\V[\cdot]\), respectively. For a set \(\mathcal{S}\), \(|\mathcal{S}|\) denotes its cardinality. The operator \(\otimes\) denotes the Kronecker product. {Finally, $\mathbf{I}_{n \times n}$ denotes the $n \times n$ identity matrix, while $\mathbf{0}_{m \times n}$ and $\mathbf{1}_{m \times n}$ denote the $m \times n$ all-zeros and all-ones matrices, respectively.}}

\vspace{-3mm}
\section{System Model, Bounds, and MLE}\label{sec:system_model}
\vspace{-1mm}
\subsection{System Model}\label{subsection:system_model}
\vspace{-1mm}
We consider an uplink scenario with a single-antenna \gls{ue} at the $i$-th unknown position $\vx_i=[x_i,y_i,z_i]^\top$ and $M$ distributed, mutually phase-synchronized single-antenna \glspl{ap} at known positions $\vp_m = [x_m, y_m, z_m]^\top$ for $m \in \{1, \dots, M\}$. The \glspl{ap} are connected through fronthaul links to a coordinating \gls{dlmf}~\cite{lmf_ref, dlmf_patent}, which collects the carrier phase measurements and performs positioning. An example scenario with three \glspl{ap} is shown in Fig.~\ref{fig:system_figure}. The \gls{ue} transmits a narrowband uplink pilot waveform $s$ with $\lvert s\rvert^2=1$ over an effective occupied bandwidth $W$, which is on the order of hundreds of kHz, with carrier wavelength $\lambda$. The bandwidth $W$ is assumed to be small such that the \gls{los} and \gls{nlos} components are not resolvable in the delay domain, and the received signal can be represented by a single carrier phase observation at each \gls{ap}. Under the two-ray multipath model~\cite{goldsmith2005wireless}, the received signal at the $m$-th \gls{ap} reads
\begin{equation}\label{eq:received_signal}
    y_{m,i} = \sqrt{P}\big(h_m^{(1)}(\vx_i) + h_m^{(2)}(\vx_i) \big) \exp{(- \jmath \phi_i)}s + w_{m,i},
\end{equation}
where $P$ is the transmit power, $h_m^{(j)}(\vx_i)$ depicts the channel of the $j$-th path, $\phi_i$ is the common \gls{cpo} between the \gls{ue} and the \gls{ap} network, and $w_{m,i} \sim \mathcal{CN}(0,\sigma_w^2)$ is additive circularly symmetric complex Gaussian noise, uncorrelated across \glspl{ap}. The channel coefficient of the $j$-th path is given by
\begin{equation}\label{eq:h_los}
    h_m^{(j)}(\vx_i) = \alpha_m^{(j)}(\vx_i) \exp{( - \jmath \theta_m^{(j)}(\vx_i) )},
\end{equation}
where $\alpha_m^{(j)}(\vx_i)$ and $\theta_m^{(j)}(\vx_i)$ are the amplitude and phase of the path, respectively. In this article, we let $h_m^{(1)}$ denote the \gls{los} component and $h_m^{(2)}$ the single-bounce \gls{nlos} component. The \gls{los} path length is $d_{m,i}^{(1)} = \lVert \vp_m - \vx_i \rVert_2$, and the \gls{nlos} path length is $d_{m,i}^{(2)} = \lVert \tilde{\vp}_m - \vx_i \rVert_2$,
where $\tilde{\vp}_m$ denotes the mirror image of $\vp_m$ with respect to the reflecting surface. The narrowband channel model~\cite{goldsmith2005wireless} assumes $d_{m,i}^{(2)} - d_{m,i}^{(1)} \ll c/W$, where $c$ is the speed of light. Our analysis focuses on ground reflection; hence, $\tilde{\vp}_m = [x_m, y_m, -z_m]^\top$ with respect to the ground plane ($z=0$). Furthermore, according to~\cite{itu_reflection_ref}, under the \gls{te} polarization assumption, the reflection coefficient of the ground reflection is given by
\begin{equation}\label{eq:gamma}
    \Gamma_m(\vx_i) = \frac{\cos(\gamma_m(\vx_i)) - \sqrt{\varepsilon_r - \sin^2(\gamma_m(\vx_i))}}{\cos(\gamma_m(\vx_i)) + \sqrt{\varepsilon_r - \sin^2(\gamma_m(\vx_i))}},
\end{equation}
where $\varepsilon_r$ is the relative permittivity of the ground material and $\gamma_m(\vx_i) = \acos(( z_i + z_m ) / d_{m,i}^{(2)})$ is the incidence angle. Now, we can express the amplitude and phase of the two paths as
\begin{subequations}
\begin{align}
    \alpha_m^{(1)}(\vx_i) &= \sqrt{G_{\textrm{tx}} G_{\textrm{rx}}}\,\lambda / (4 \pi d_{m,i}^{(1)}), \\
    \theta_m^{(1)}(\vx_i) &= 2 \pi d_{m,i}^{(1)} / \lambda, \\
    \alpha_m^{(2)}(\vx_i) &= \lvert \Gamma_m(\vx_i) \rvert \sqrt{G_{\textrm{tx}} G_{\textrm{rx}}}\,\lambda / (4 \pi d_{m,i}^{(2)}), \\
    \theta_m^{(2)}(\vx_i) &= 2 \pi d_{m,i}^{(2)} / \lambda - \angle \Gamma_m(\vx_i),
\end{align}
\end{subequations}
where $G_{\textrm{tx}}$ and $G_{\textrm{rx}}$ denote the transmit and receive antenna gains and for simplicity, we assume $G_{\textrm{tx}} = G_{\textrm{rx}} = 1$.\footnote{The isotropic gain assumption is adopted for analytical simplicity. In practice, directional transmit and receive antenna patterns, as well as the orientations of the \gls{ue} and \glspl{ap}, can introduce path-dependent gains for the \gls{los} and \gls{nlos} components. Such gains would change the relative amplitudes of the two phasors in \eqref{eq:theta_eff_closedform}, and hence may affect both the effective carrier phase $\theta_m(\vx_i)$ and the phase noise term in the carrier phase measurement model in \eqref{eq:carrier_phase_measurements}, whose variance is given by $\sigma_{m,i}^2$ in \eqref{eq:phase_noise_variance}. Incorporating orientation-dependent antenna patterns is an important topic for future work.}

The functional form of the received signal's amplitude and phase is not apparent from \eqref{eq:received_signal}. Therefore, we re-express the received signal using the phasor representation as
\begin{equation}\label{eq:alt_received_signal}
    y_{m,i} = \sqrt{P} \alpha_{m}(\vx_i) \exp{\left(-\jmath (\theta_{m}(\vx_i) + \phi_i)\right)}s + w_{m,i},
\end{equation}
where the effective amplitude and phase can be expressed as
\begin{subequations}
\begin{align}
    \alpha_{m}(\vx_i) &= \sqrt{ \tilde{\alpha}_1^2 + \tilde{\alpha}_2^2
+ 2 \tilde{\alpha}_1 \tilde{\alpha}_2
\cos \left( \tilde{\theta}_1 - \tilde{\theta}_2 \right) }, \label{eq:alpha_eff_closedform} \\
 \theta_m(\vx_i) &= \atan2 \left( \sum \nolimits_{j} \tilde{\alpha}_j \sin(\tilde{\theta}_j),\sum \nolimits_{j} \tilde{\alpha}_j \cos(\tilde{\theta}_j) \right). \label{eq:theta_eff_closedform}
\end{align}
\end{subequations}
In the above, the amplitudes and phases of the two paths ($j \in \{1, 2\}$) have been denoted using $\tilde{\alpha}_j \triangleq \alpha_m^{(j)}(\vx_i)$ and $\tilde{\theta}_j \triangleq \theta_m^{(j)}(\vx_i)$ for brevity, and atan2$(y, x)$ denotes the four-quadrant arc tangent function. From the received signal $y_{m,i}$ given in \eqref{eq:alt_received_signal}, each \gls{ap} obtains a wrapped carrier phase measurement $r_{m,i} = -\angle y_{m,i} \in [-\pi, \pi)$ which can be mathematically expressed as
\begin{equation}\label{eq:carrier_phase_measurements}
    r_{m,i} = \textrm{wrap}_{[-\pi,\pi)}\left(\theta_m(\vx_i) + \phi_i + n_{m,i}\right),
\end{equation}
where $n_{m,i}$ is the noise of the carrier phase measurements and
\begin{equation}
    \textrm{wrap}_{[a,b)}(\zeta) = \textrm{mod}(\zeta-a,b-a)+a
\end{equation}
is the wrap-around function that wraps the angle $\zeta$ in between $[a,b)$. 
{The noise is modeled as $n_{m,i} \sim \mathcal{N}(0,\sigma_{m,i}^2)$, where $\sigma_{m,i}^2$ is obtained as the corresponding diagonal entry of the inverse of the carrier phase \gls{fim} block in \eqref{eq:FIM_block_partition}, expressed as}
\begin{equation}\label{eq:phase_noise_variance}
\sigma_{m,i}^2 =
\frac{\sigma_w^2}{2P\alpha_m^2(\vx_i)},
\end{equation}
where $\alpha_m(\vx_i)$ is the effective amplitude in \eqref{eq:alpha_eff_closedform}. In this article, we assume that $\sigma_{m,i}^2$ is given by the channel estimation routine that computes the phase, and the relation to $\vx_i$ is not exploited by the developed estimators.

\vspace{-2mm}
\subsection{Problem Formulation}
Rather than relying on the full received signal $\vy_i=[y_{1,i},\ldots,y_{M,i} ]^\top$, we consider the carrier phase measurement vector $\vr_i=[r_{1,i},\ldots,r_{M,i} ]^\top$ as the main input for positioning. The main objective is to estimate the unknown \gls{ue} position $\vx_i$ from the phase observation vector $\vr_i$ using a low-complexity and high-accuracy single-snapshot positioning framework. 

\textls[-2]{The resulting positioning problem is challenging for several reasons. First, the carrier phase measurements in \eqref{eq:carrier_phase_measurements} are available only in wrapped form, which leads to the well-known integer ambiguity problem~\cite{10232971}. Second, the unknown common phase offset $\phi_i$ further complicates the estimation task, since it couples all measurements across the \gls{ap} network. Third, under the considered two-ray propagation model, the effective phase depends on the \gls{ue} position through a highly nonlinear mapping determined by the propagation distances and the reflection coefficient. The reflection coefficient in \eqref{eq:gamma} depends on the relative permittivity $\varepsilon_r$ and the polarization mode. In this work, the reflecting surface is assumed to be planar and homogeneous, so that $\varepsilon_r$ is constant over the considered area. However, $\varepsilon_r$ and the polarization mode are not assumed to be known by the proposed learning-based positioning method; they are only required by the analytical model used for the \gls{peb} derivation in Section~\ref{subsection:performance_bounds} and the model-based \gls{mle} in Section~\ref{subsection:mle}. The resulting localization problem is highly non-convex, so that the associated maximum-likelihood objective can contain multiple local peaks and yield ambiguous location estimates~\cite{dey2025polo}. While \gls{egs} may recover the correct position by densely sampling the search region, such an approach suffers from high complexity~\cite{11275391}.}

\vspace{-3mm}
\subsection{Performance Bounds}\label{subsection:performance_bounds}
Let $\veta = [\phi, \alpha_1, \ldots, \alpha_M, \theta_1, \ldots, \theta_M]^\top$ denote the channel parameter vector of the $M$ received signals, and the dependence on $i$ and $\vx_i$ is omitted here for brevity. The \gls{fim} of the unknown parameters is given by~\cite[Section~15.7]{kay_est_theory}
\begin{equation}\label{eq:channel_fim_main}
[\vJ(\veta)]_{kl} = \frac{2}{\sigma_w^2} \sum_{m=1}^{M} \Re \left \{
\frac{\partial \bar{y}_m^*}{\partial \veta_k}
\frac{\partial \bar{y}_m}{\partial \veta_l} \right \}, \:
k,l \in \{1,\ldots,2M+1\},
\end{equation}
where $\bar{y}_m$ denotes the noise-free signal component of $y_m$ in \eqref{eq:alt_received_signal}. {Let $\valpha=[\alpha_1,\ldots,\alpha_M]^\top$ and $\vtheta=[\theta_1,\ldots,\theta_M]^\top$. Then the \gls{fim} can be block-partitioned as follows
\begin{equation}
\vJ(\veta)=
\begin{bmatrix}
J_{\phi\phi}          & \vJ_{\phi\valpha}          & \vJ_{\phi\vtheta}\\
\vJ_{\phi \valpha}^\top     & \vJ_{\valpha\valpha}       & \vJ_{\valpha\vtheta}\\
\vJ_{\phi \vtheta}^\top     & \vJ_{\valpha \vtheta}^\top       & \vJ_{\vtheta\vtheta}
\end{bmatrix}.
\label{eq:FIM_block_partition}
\end{equation}
The block diagonal elements of the \gls{fim} are given by
$J_{\phi\phi} = \frac{2P}{\sigma_w^2} \sum_{m=1}^M \alpha_m^2$, 
$\vJ_{\valpha\valpha} = \frac{2P}{\sigma_w^2}\vI_{M \times M}$ and
$\vJ_{\vtheta\vtheta}  = \frac{2P}{\sigma_w^2} \vLambda$ 
in which $\vLambda =  \diag([\alpha_1^2,\ldots,\alpha_M^2])$. Respectively, the cross-elements of the \gls{fim} are given by
$\vJ_{\phi\valpha} = \mathbf{0}_{1 \times M}$,
$\vJ_{\valpha\vtheta}  = \mathbf{0}_{M \times M}$ and
$\vJ_{\phi\vtheta}  = \frac{2P}{\sigma_w^2} \valpha_{\text{sq}}^{\top}$ 
in which $\valpha_{\mathrm{sq}} = [\alpha_1^2,\ldots,\alpha_M^2]^\top$. The \gls{efim} corresponding to $\vtheta$ is derived by partitioning $\veta$ into the parameters of interest (i.e., $\vtheta$) and the nuisance parameters (i.e., $\vnu = [\phi, \valpha^\top]^\top$). Accordingly, the \gls{fim} is block-partitioned as
$
    \vJ(\veta) = \left[ \begin{smallmatrix}
        \vJ_{\vnu \vnu} & \vJ_{\vnu \vtheta} \\
        \vJ_{\vnu \vtheta}^\top & \vJ_{\vtheta \vtheta} \\
    \end{smallmatrix} \right]
$
and} 
the \gls{efim} can be obtained using the Schur complement as 
\begin{subequations}
\begin{align}
    \vJ(\vtheta) &= \vJ_{\vtheta\vtheta} - \vJ_{\vnu\vtheta}^\top \vJ_{\vnu\vnu}^{-1}\vJ_{\vnu\vtheta}, \label{eq:efim_phase1} \\
    &= \frac{2P}{\sigma_w^2} \left( \vLambda - \frac{\valpha_{\mathrm{sq}} (\valpha_{\mathrm{sq}})^\top}{ \sum_{m=1}^{M} \alpha_m^2} \right). \label{eq:efim_phase2}
\end{align}
\end{subequations}
Equivalently, we could arrive at the same expression of \gls{efim} in \eqref{eq:efim_phase2} using the model in \eqref{eq:carrier_phase_measurements} and the variance of the phase measurements in \eqref{eq:phase_noise_variance}. It is important to note that the \gls{fim} and \gls{efim} above are rank-deficient because the unknown phases and \gls{cpo} are coupled. However, the \gls{fim} of the \gls{ue} state can still have full rank as long as \glspl{ap} are phase-synchronized, and can be computed using transformation of variables as
\begin{equation}\label{eq:ue_fim}
 \vJ(\vx) = [\vH_\vx(\vtheta)]^\top \vJ(\vtheta) \vH_\vx(\vtheta),   
\end{equation}
where the Jacobian is $[\vH_\vx(\vtheta)]_{m,j} = \partial [\vtheta]_m / \partial [\vx]_j \in \mathbb{R}^{M \times \textrm{dim}(\vx)}$, {with the corresponding detailed derivations provided in the Appendix.} Note that, through the reflection coefficient in~\eqref{eq:gamma} entering both $\vJ(\vtheta)$ and the Jacobian $\vH_\vx(\vtheta)$, the \gls{fim} in~\eqref{eq:ue_fim} presumes perfect knowledge of the relative permittivity $\varepsilon_r$ and the polarization mode. Finally, the \glsentrylong{peb} is computed from $ \vJ(\vx) $ using $\text{PEB} = \sqrt{\trace(\vJ(\vx)^{-1})}$. 



\vspace{-2.5mm}
\subsection{Maximum Likelihood Estimator (MLE)}\label{subsection:mle}
\textls[-3]{Since the measurement noise in \eqref{eq:carrier_phase_measurements} is zero-mean Gaussian, the variance in \eqref{eq:phase_noise_variance} is small, and the von Mises distribution can be approximated using a Gaussian~\cite{abdi2002}. Furthermore, the measurements are assumed to be independent, thus the likelihood function can be expressed as}
\begin{equation}
    p(\vr_i \mid \vx_i, \phi_i) = \prod_{m=1}^M  \N(r_{m,i}; \mu_m(\vx_i,\phi_i), \sigma_{m,i}^2),
\end{equation}
in which $\mu_m(\vx_i,\phi_i) = \theta_m(\vx_i) + \phi_i$ and $\sigma_{m,i}^2$ is defined in \eqref{eq:phase_noise_variance}. In order to express the cost function that we aim to minimize only in terms of the unknown \gls{ue} position, we follow~\cite{dey2025polo} and compress away the unknown $\phi_i$. This is achieved by differentiating the log-likelihood function with respect to $\phi_i$, equating to zero and solving for $\phi_i$ to obtain
\begin{equation}\label{eq_cfo_estimate}
    \hat{\phi}(\vx_i) =  \frac{\sum_{m=1}^M \varepsilon_m(\vx_i)/\sigma_{m,i}^2}{\sum_{m=1}^M 1/\sigma_{m,i}^2},
\end{equation}
where $\varepsilon_m(\vx_i)$ is the (wrap-corrected)
phase residual at AP~$m$, with the wrap-around handling formally specified
in~\eqref{eq:residual_correction} below.
When the \gls{cpo} is close to $\pm\pi$, the uncorrected residuals
$\tilde{\varepsilon}_m(\vx_i) = \textrm{wrap}_{[-\pi,\pi)}(r_{m,i} - \theta_m(\vx_i))$
may cross the wrapping boundary, and therefore the wrap-around effect must be handled carefully. One approach is to correct for the residuals as 
\begin{equation}\label{eq:residual_correction}
\varepsilon_m(\vx_i) = \begin{cases} 
\tilde{\varepsilon}_m(\vx_i) + 2\pi & \text{if} \quad \lvert \cI_+ \rvert > \lvert \cI_- \rvert \wedge m \in \cI_-, \\
\tilde{\varepsilon}_m(\vx_i) - 2\pi & \text{if} \quad \lvert \cI_+ \rvert < \lvert \cI_- \rvert \wedge m \in \cI_+, \\
\tilde{\varepsilon}_m(\vx_i) & \text{otherwise},
\end{cases}   
\end{equation}
where $\cI_+ = \{m \in \cM \mid \tilde{\varepsilon}_m(\vx_i) > \pi/2 \}$, $\cI_- = \{m \in \cM \mid \tilde{\varepsilon}_m(\vx_i) < -\pi/2 \}$, and $\cM = \{1,\ldots,M\}$. Plugging the \gls{cpo} estimate above into the likelihood function (i.e., $\mu_m(\vx_i,\hat{\phi}_i) = \theta_m(\vx_i) + \hat{\phi}(\vx_i)$), the localization problem now only depends on $\vx_i$ and the \gls{mle} for the \gls{ue} position can be obtained by minimizing the negative log-likelihood, expressed as
\begin{align}
    \hat{\vx}_i &= \underset{\vx_i}{\argmin} \, C(\vx_i), \quad \text{where} \label{eq:mle} \\
    C(\vx_i) &= \sum_{m=1}^M  \frac{1}{ \sigma_{m,i}^2}[\textrm{wrap}_{[-\pi,\pi)}(r_{m,i} - \mu_m(\vx_i,\hat{\phi}_i))]^2.
\end{align}
The defined problem can be solved using nonlinear optimization~\cite{boyd2004}. However, the highly non-convex cost function makes the optimization algorithm very sensitive to initialization~\cite{10851402} and therefore, we adopt a two-step procedure to solve the problem~\cite{dey2025polo}. In the first step, we compute the \gls{ls} estimate using \gls{egs} over a predefined search set of candidate \gls{ue} positions $\cX$, mathematically defined as
\begin{equation}\label{eq:egs}
    \hat{\vx}_i^{(0)} = \underset{\vx_i \in \cX}{\argmin} \, \tilde{C}(\vx_i),
\end{equation}
where $\tilde{C}(\vx_i) = \sum_{m=1}^M [\textrm{wrap}_{[-\pi,\pi)}(r_{m,i} - \mu_m(\vx_i,\hat{\phi}_i))]^2$ is the \gls{ls} cost function.\footnote{\gls{egs} introduces a non-negligible discretization error for which the variance is given by $\sigma_{\textrm{EGS}}^2 = (\delta_x \pi / \lambda)^2$ in which $\delta_x$ denotes the grid spacing. \gls{egs} can also be calculated using $C(\vx_i)$ and the robustness of the algorithm can be improved by inflating the variance to $\sigma_{m,i}^2 + \sigma_{\textrm{EGS}}^2$ in order to account for the discretization error. Since $ \sigma_{\textrm{EGS}}^2 \gg \sigma_{m,i}^2$, the inflated variance values are nearly identical for every $\vx_i \in \cX$ such that \gls{egs} computed using \gls{ls} and modified \gls{mle} costs yield very similar performance. Since \gls{ls} has a lower computational overhead, we opted to use the \gls{ls} cost in this article.}
\textls[-6]{In the second step, the final estimate $\hat{\vx}_i$ is obtained by solving \eqref{eq:mle} using \gls{gd} and the optimization algorithm is initialized using $\hat{\vx}_i^{(0)}$. As with the \gls{peb}, the \gls{mle} in~\eqref{eq:mle} and \gls{ls} in~\eqref{eq:egs} presume perfect knowledge of relative permittivity $\varepsilon_r$ and the polarization mode through the analytical phase model $\theta_m(\vx_i)$ in~\eqref{eq:theta_eff_closedform}. {We note that the derived MLE for D-MIMO phase-only positioning under dominant multipath has not been presented earlier in the literature.}}

\vspace{-2mm}
\section{Gaussian Process Carrier Phase Model}\label{sec:gaussian_process}
\begin{figure*}[!t]
    \centering
    \vspace{-3mm}
    \includegraphics[width=0.95\textwidth]{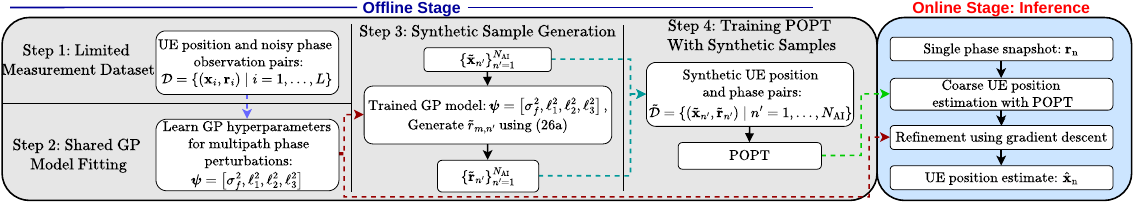}
    \caption{Offline-online pipeline of the proposed overall positioning framework. In the offline stage, a limited measurement dataset with $L$ samples is used to learn a shared \gls{gp} model, generate $N_\mathrm{AI}$ synthetic samples, and train \gls{popt} on the resulting synthetic training set. In the online stage, a single phase snapshot is used to obtain a coarse \gls{ue} position estimate, which is further refined through gradient descent.}
    \vspace{-4mm}
    \label{fig:overall_framework}
\end{figure*}

Figure~\ref{fig:overall_framework} provides an overview of the proposed overall offline--online positioning framework. In this section, we focus mainly on the offline \gls{gp}-learning stage, where a limited set of measurement data is used to learn a shared model for the carrier phase and to support subsequent synthetic data generation. Instead of relying on the theoretical carrier phase model presented in Section~\ref{sec:system_model}, which requires knowing the polarization type and relative permittivity of the material, we resort to \gls{gp} regression to model the carrier phase measurements. \glspl{gp} provide a Bayesian non-parametric approach for data-driven modeling of smooth functions~\cite{Rasmussen+Williams:2006}, and the proposed \gls{gp} model bridges the theoretical propagation model in Section~\ref{subsection:system_model} and the learning-based positioning method developed in Section~\ref{sec:positioning}.

\vspace{-2mm}
\subsection{Proposed GP Model}\label{sec:proposed_gp_model}
\begin{figure*}[!t]
  \vspace{-7mm}
  \centering
  \subfloat[NLoS phase perturbation in \eqref{eq:nlos_phase_perturbation}]{%
    \includegraphics[width=0.32\textwidth]{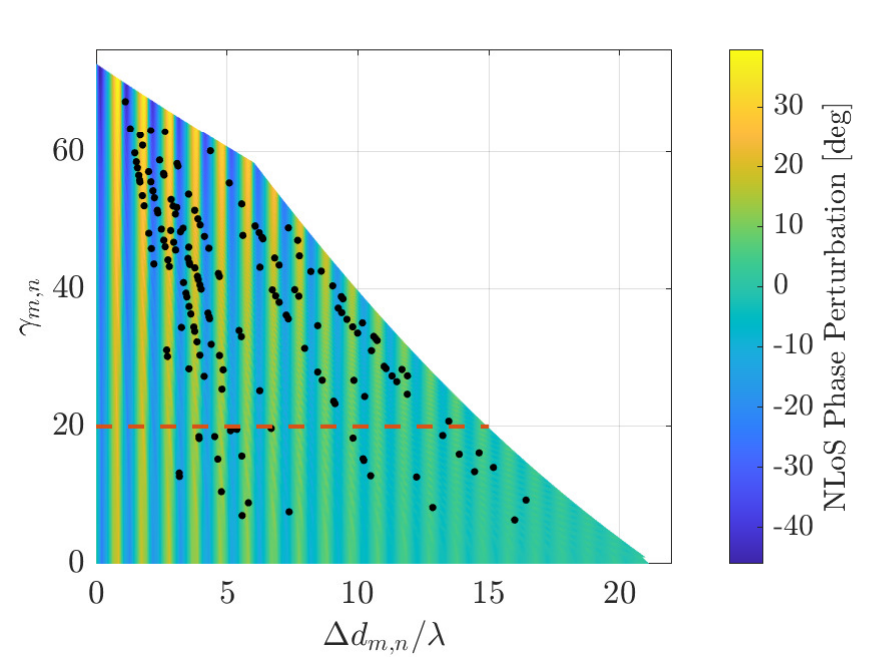}%
    \label{fig:gp_example1}%
  }\hfill
  \subfloat[True and estimated phase perturbations]{%
    \includegraphics[width=0.32\textwidth]{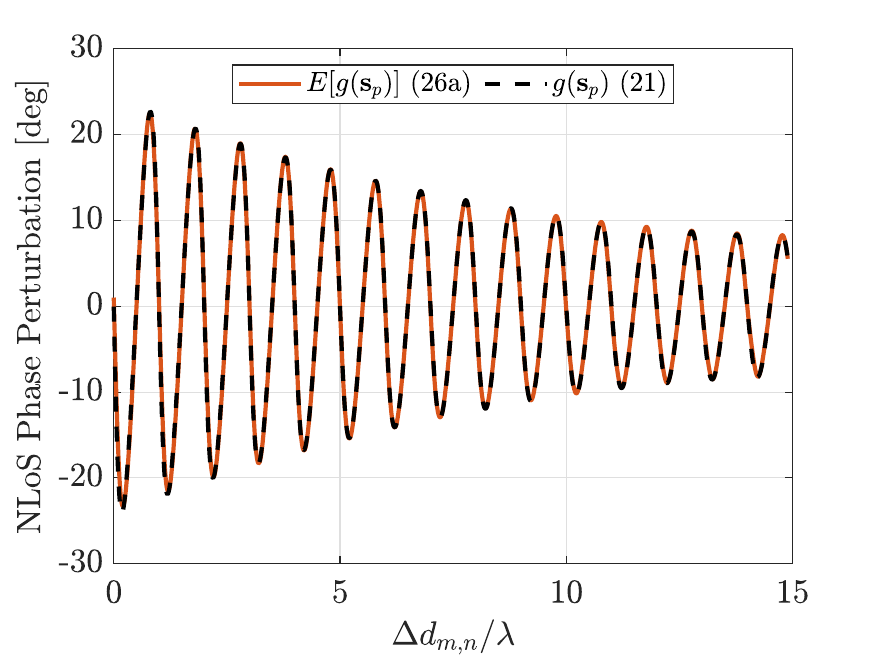}%
    \label{fig:gp_example2}%
  }\hfill
  \subfloat[True and estimated carrier phase]{%
    \includegraphics[width=0.32\textwidth]{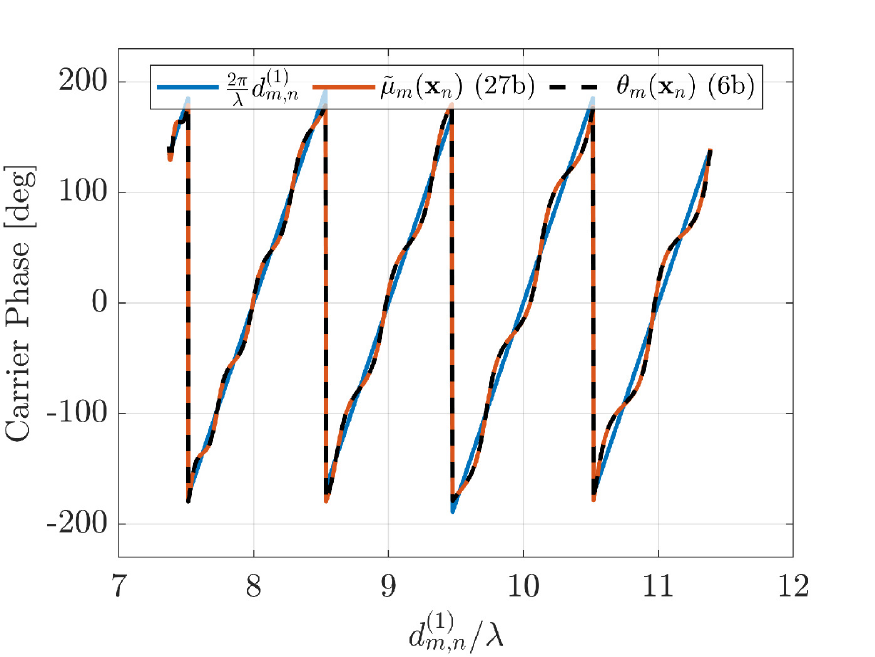}%
    \label{fig:gp_example3}%
  }
  \vspace{2mm}
  \caption{An example illustration of the proposed GP model. The experimental setting is described in Section~\ref{sec:num_results} with $M=17$ \glspl{ap}, $L=10$ \gls{ue} locations used for training, and the transmit power is set to $P=-10\,\textrm{dBm}$. The \gls{nlos} phase perturbations across different incidence angles and path length differences are visualized in (a) in which quasi-periodicity is clearly visible. The black markers visualize the GP training points ($M \times L = 170$) and the red line indicates a slice of the model that is more closely visualized in (b) and (c). The predicted and theoretical phase perturbations are visualized in (b) as a function of $\Delta d_{m,n}$ for one \gls{ap} and fixed incidence angle $\gamma_{m,n}=20~\text{deg}$. The red line illustrates the GP predictions and the dashed black line visualizes the theoretical phase perturbation. In (c), the \gls{los}-only carrier phase is shown using the blue line, the carrier phase predicted by the proposed model is illustrated with the red line, and the carrier phase of the theoretical model in  \eqref{eq:theta_eff_closedform} is shown with the black dashed line. The CPO is set to zero in the example for visual clarity.}
    \vspace{-2mm}
  \label{fig:gp_example}
\end{figure*}

In this article, we use \glspl{gp} to model the carrier phase measurements given in \eqref{eq:carrier_phase_measurements}. Hence, for the $m$-th \gls{ap} and $i$-th \gls{ue} location pair, the following model is proposed 
\begin{equation}\label{eq:proposed_gp_model}
    r_{m,i} = \textrm{wrap}_{[-\pi,\pi)}\left(\tfrac{2 \pi}{\lambda} d_{m,i}^{(1)} + \phi_i + g(\vs_q) + n_{m,i}\right),
\end{equation}
where $\vs_q = [\vx_i^\top, \, \vp_m^\top]^\top$ and $q = (i-1) M + m$ are used for notational convenience. In \eqref{eq:proposed_gp_model}, the first two terms inside the wrapping function capture the carrier phase of the \gls{los} component and \gls{cpo}, the third term models the phase perturbations induced by the \gls{nlos} signal, and the last term is input-dependent measurement noise given in \eqref{eq:phase_noise_variance}. The phase perturbation of the \gls{nlos} signal is given by
\begin{equation}\label{eq:nlos_phase_perturbation}
    g(\vs_q) = \textrm{wrap}_{[-\pi,\pi)}\left(\theta_m(\vx_i) - \tfrac{2 \pi}{\lambda} d_{m,i}^{(1)} \right),
\end{equation}
\textls[-3]{and since the relative permittivity and polarization type are assumed unknown, we assign a zero-mean \gls{gp} prior to the unknown function, that is, $g(\vs_q) \sim \GP(0,k(\vs_q,\vs_{q'}))$ with covariance $k(\vs_q,\vs_{q'})$~\cite{Rasmussen+Williams:2006}. The \gls{nlos} phase perturbation in \eqref{eq:nlos_phase_perturbation} is illustrated in Fig.~\ref{fig:gp_example1}. The proposed model assumes the following:
\begin{enumerate*}[label=(\roman*)]
  \item the normal to the reflection surface is known; and
  \item the dielectric properties of the surface material are homogeneous.
\end{enumerate*}
These assumptions allow us to design the covariance function so that it takes advantage of the periodicity of the phase measurements. In addition, we can assign a single shared \gls{gp} prior to the \glspl{ap} facilitating training with sparse data. If the above assumptions do not hold, a \gls{gp} model is still a viable option, but in this case every \gls{ap} would have to be assigned its own prior.}


The covariance function is a measure of the similarity between different function values. Its design and choice are key when employing \gls{gp} models, and mathematical understanding of the problem can be leveraged to impose strong theoretical constraints on top of observational ones. The first choice in designing the covariance function is the regressor, and an obvious candidate is the \gls{ue} location. However, this choice does not explain the periodicity of the data very well and a better alternative is to use two regressors\textemdash the difference between the \gls{nlos} and \gls{los} path distances, $\Delta d_{m,i} = d_{m,i}^{(2)} - d_{m,i}^{(1)}$ and the incidence angle, $\gamma_{m,i} = \acos(( [\vx_i]_3 + [\vp_m]_3 ) / d_{m,i}^{(2)})$. 
The second important choice is the covariance function itself, and since the phase perturbation of the \gls{nlos} signal is not exactly periodic as shown in Fig.~\ref{fig:gp_example1}, it is desirable to allow for quasi-periodic variation, where the shape of the periodic effect can change over time~\cite{solin2014}. A common way to construct quasi-periodic covariances is to take the product of a periodic covariance function with another covariance function, allowing the covariance to decay away from the exact periodicity~\cite{Rasmussen+Williams:2006,solin2014}. In this article, we construct the covariance function as a product of three covariance functions, expressed as
\begin{multline}\label{eq:quasi_periodic}
        k(\vs_q, \vs_{q'}) = \sigma_f^2 
    k_{\textrm{pe}}(\Delta d_{m,i}, \Delta d_{m',i'} \mid \ell_1^2) \\
    \times k_{\textrm{se}}(\Delta d_{m,i}, \Delta d_{m',i'}  \mid \ell_2^2) 
    \, k_{\textrm{se}}(\gamma_{m,i},\gamma_{m',i'} \mid \ell_3^2),
\end{multline}
where $\sigma_f^2 $, $\ell_1^2$, $\ell_2^2$ and $\ell_3^2$ represent the magnitude scale and three characteristic length-scales, respectively. In the above, the widely utilized periodic and squared exponential covariance functions are defined as~\cite{Rasmussen+Williams:2006}
\begin{subequations}
   \begin{align}
    k_{\textrm{pe}}(x,x' \mid \ell^2) &= \exp\left(-\tfrac{2}{\ell^2} \sin^2\left( \omega_0 \tfrac{x - x'}{2} \right)\right), \\
    k_{\textrm{se}}(x,x' \mid \ell^2) &= \exp\left(-\tfrac{1}{2 \ell^2} \lVert x - x' \rVert^2\right),
\end{align} 
\end{subequations}
\textls[-9]{where $\omega_0 = 2 \pi /\lambda$ is the known frequency scale parameter defined by the wavelength. The proposed model implies that the \gls{nlos} signal causes periodic fluctuations in the carrier phase as a function of $\Delta d_{m,i}$ and with the period defined by the wavelength. In addition, the \gls{nlos} perturbations are quasi-periodic, which allows the shape of the periodic effect to change as a function of $\Delta d_{m,i}$ and $\gamma_{m,i}$.} 


\vspace{-4mm}
\subsection{Training the GP Model}\label{sec:gp_training}
Step 1 in Fig.~\ref{fig:overall_framework} corresponds to the acquisition of a training dataset in $L$ \gls{ue} locations, given by $\mathcal{D} = \{(\vx_i, \vr_i) \mid i = 1,\ldots, L \}$ in which $\vr_i = [r_{1,i}, \, \ldots, \, r_{M,i}]^\top$. The subsequent learning of the shared \gls{gp} model is then performed as illustrated in Step 2 of Fig.~\ref{fig:overall_framework}. Consequently, the unknown \glspl{cpo} are $\vphi = [\phi_1, \, \phi_2, \, \ldots, \, \phi_L]^\top$, the hyperparameters of the proposed \gls{gp} model are $\vpsi = [\sigma_f^2, \, \ell_1^2, \, \ell_2^2, \, \ell_3^2]^\top$, and a straightforward way to learn them is by maximizing the marginal likelihood function~\cite{Rasmussen+Williams:2006}. The log marginal likelihood, $\cL_{\vr} \triangleq \log p(\vr \mid \vphi, \vpsi)$, and its derivatives now read
\begin{subequations}
\begin{align}
    \cL_{\vr} &= -\frac{1}{2} \left(\vupsilon^\top \vK_{\vr}^{-1} \vupsilon  + \log \lvert \vK_{\vr} \rvert + ML \log(2\pi)\right), \label{eq:log_marginal_likelihood} \\
     \frac{\partial \cL_{\vr}}{\partial \vphi} &= (\vI_{L \times L} \otimes \mathbf{1}_{M \times 1})^\top \vK_{\vr}^{-1} \vupsilon, \\
     \frac{\partial \cL_{\vr}}{\partial [\vpsi]_j} &= \frac{1}{2} \vupsilon^\top \vK_{\vr}^{-1} \frac{\partial \vK_{\vr}}{\partial [\vpsi]_j}\vK_{\vr}^{-1}  \vupsilon - \frac{1}{2} \trace\left( \vK_{\vr}^{-1} \frac{\partial \vK_{\vr}}{\partial [\vpsi]_j} \right), 
\end{align}
\end{subequations}
where $\vupsilon \in \mathbb{R}^{ML \times 1}$ is the residual vector and $\vK_{\vr} \in \mathbb{R}^{ML \times ML}$ the covariance matrix. The elements of $\vupsilon$ and $\vK_{\vr}$ are
\begin{subequations}
\begin{align}
     [\vupsilon]_q &= \textrm{wrap}_{[-\pi,\pi)}(r_{m,i} - \tfrac{2 \pi}{\lambda} d_{m,i}^{(1)} - \phi_i), \\
    [\vK_{\vr}]_{q,q'} &=  k(\vs_q, \vs_{q'}) + \delta_{q,q'} \sigma_{m,i}^2,   
\end{align} 
\end{subequations}
where $\delta_{q,q'}$ is the Kronecker delta, equal to one if $q = q'$ and zero otherwise.
Once the marginal likelihood and its derivatives are available, the model parameters can be trained using, for example, a gradient-based optimizer.

\vspace{-5mm}
\subsection{Gaussian Process Regression}\label{sec:gp_regression}
In the following, we use \gls{ue} index $n$ and the shorthand notation $\vs_p = [\vx_n^\top, \, \vp_m^\top]^\top$ in which $p = (n-1)M+m$ to make a clear distinction between training and test data. \gls{gp} regression is usually formulated as predicting the unknown function value $g(\vs_p)$ associated with a known test point $\vx_n$ given a training dataset $\mathcal{D}$. This can be achieved by finding the predictive density $p(g(\vs_p) \mid \mathcal{D}) =  \N(g(\vs_p) \mid \E[g(\vs_p)], \V[g(\vs_p)])$, with the mean and variance given by~\cite{Rasmussen+Williams:2006}
\begin{subequations}
    \begin{align}
        \E[g(\vs_p)]
            & = \vk_p^\top \vK_\vr^{-1}\vupsilon, \label{eq:gp_predictive_mean} \\
        \V[g(\vs_p)]
            & = k(\vs_p,\vs_p) - \vk_p^\top \vK_\vr^{-1} \vk_p, \label{eq:gp_predictive_variance}
    \end{align}
\end{subequations}
where $\vk_p = [k(\vs_p, \vs_1), \, \dots, \, k(\vs_p, \vs_{ML})]^\top \in \mathbb{R}^{ML \times 1}$ is the cross-covariance between the test and training data. Using \eqref{eq:gp_predictive_mean}, \eqref{eq:gp_predictive_variance} and modeling the noise as $\tilde{n}_{m,n} \sim \mathcal{N}(0,\tilde{\sigma}_{m,n}^2)$, the carrier phase measurements of the proposed model for the $m$-th \gls{ap} and $n$-th \gls{ue} location can now be expressed as
\begin{subequations}
\begin{align}
    \tilde{r}_{m,n} &= \textrm{wrap}_{[-\pi,\pi)}\left( \tilde{\mu}_m(\vx_n) + \tilde{n}_{m,n} \right),  \label{eq:gp_proposed_phase} \\
    \tilde{\mu}_m(\vx_n) &= \tfrac{2 \pi}{\lambda} d_{m,n}^{(1)} + \phi_n + \E[g(\vs_{p})] , \label{eq:gp_proposed_mean} \\
    \tilde{\sigma}_m^2(\vx_n) &= \sigma_{m,n}^2 + \V[g(\vs_{p})]. \label{eq:gp_proposed_variance}
\end{align}
\end{subequations}
The proposed model is illustrated in Fig.~\ref{fig:gp_example} trained using $L=10$ \gls{ue} locations. The quasi-periodicity of the \gls{nlos} perturbations is accurately captured by the \gls{gp} model and the estimation error of the proposed model is below one degree in the example.

We utilize the above model in two different ways. First, the trained model allows us to generate arbitrarily many training samples for the \gls{dl}-based positioning approach as illustrated in Step 3 of Fig.~\ref{fig:overall_framework}. Let us denote the synthetic training data as $\tilde{\mathcal{D}} = \{(\tilde{\vx}_{n'}, \tilde{\vr}_{n'}) \mid n' = 1,\ldots, N_\textrm{AI} \}$ in which $\tilde{\vr}_{n'} = [\tilde{r}_{1,n'}, \, \ldots, \, \tilde{r}_{M,n'}]^\top$. Using the trained \gls{gp} model, the synthetic phase measurements are generated using \eqref{eq:gp_proposed_phase}. For each \gls{ue} location $\tilde{\vx}_{n'}$, an independent \gls{cpo} $\tilde{\phi}_{n'}$ is drawn uniformly from $[-\pi,\pi)$ and used in \eqref{eq:gp_proposed_mean}, i.e., by setting $\vx_n=\tilde{\vx}_{n'}$ and $\phi_n=\tilde{\phi}_{n'}$ when generating $\tilde{r}_{m,n'}$. This prevents the proposed positioning model from learning a fixed \gls{cpo} pattern and promotes invariance to the unknown \gls{cpo}. Second, the coarse position estimates of \gls{popt} are refined using the \gls{gp} model in combination with \gls{gd} as shown in the online stage of Fig.~\ref{fig:overall_framework}. This is achieved by replacing the mean and variance of the proposed \gls{mle} in Section~\ref{subsection:mle} with the mean and variance given in \eqref{eq:gp_proposed_mean} and \eqref{eq:gp_proposed_variance}, respectively. Using \gls{gd} with the \gls{gp} model, the \gls{cpo} is estimated as described in Section~\ref{subsection:mle}.

\vspace{-2mm}
\section{Proposed Positioning Approach}\label{sec:positioning}
\textls[-11]{In this section, we present the proposed positioning approach for phase-only positioning. We introduce \gls{popt}, an encoder-only transformer architecture that estimates the \gls{ue} position directly from the carrier phase observations. In the online stage of the overall framework shown in Fig.~\ref{fig:overall_framework}, a single carrier phase snapshot is processed by \gls{popt} to obtain a coarse \gls{ue} position estimate, which is subsequently refined using \gls{gd} as described in Section~\ref{subsection:mle}. Finally, we describe two benchmark methods adopted to evaluate the performance of the proposed \gls{popt}.}

\vspace{-5mm}
\subsection{Proposed POPT Architecture}\label{subsection:popt}
\textls[-3]{The phase measurements in \eqref{eq:carrier_phase_measurements} are ambiguous due to wrapping and the unknown \gls{cpo}. However, the phase relationships across the \glspl{ap} contain rich spatial information about the \gls{ue} position. Therefore, an effective learning-based positioning model should be able to jointly process all \gls{ap}-wise observations and capture their interdependencies. Transformers~\cite{attention_is_all_you_need} are particularly well suited to this requirement, and have recently been shown to outperform conventional \glspl{nn} in positioning tasks~\cite{tran_outperforms1, tran_outperforms2, tran_outperforms3}. Their underlying self-attention mechanism assigns input-dependent weights across the elements of the input sequence, enabling the model to adaptively capture interactions between any pair of elements and focus on the most task-relevant information. Inspired by \gls{vit}~\cite{vit}, we propose \gls{popt}, an encoder-only transformer tailored for \gls{ue} positioning that {processes the \gls{ap}-wise phase measurements as a token sequence and associates each token with its \gls{ap} through a learnable \gls{ap} embedding.}}

\begin{figure}[t!]
    \centering
    \vspace{-2mm}
  \includegraphics[width=0.4\textwidth]{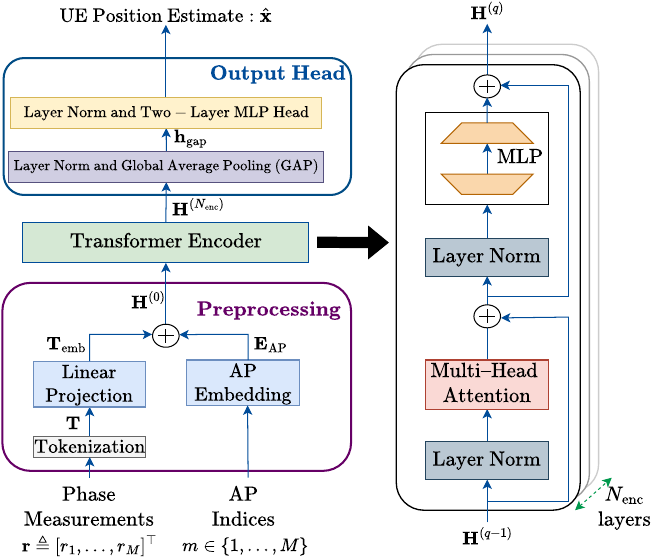}
  \caption{{Architecture of the proposed Phase-Only Positioning Transformer (POPT), which uses an encoder-only self-attention structure with learnable \gls{ap} embeddings to learn inter-\gls{ap} phase relationships for 3D \gls{ue} positioning.}}
  \vspace{2mm}
  \label{fig:popt}
\end{figure}

\textls[-3]{An overview of the proposed positioning algorithm based on \gls{popt} is shown in Fig.~\ref{fig:popt}. As detailed in the following subsections, the proposed method consists of three components: Preprocessing, Transformer Encoder, and the Output Head for \gls{ue} position regression.}

\subsubsection{Preprocessing}
\textls[-3]{For each phase measurement $r_m$, we define a token
$\vt_m \triangleq [\sin(r_m),\ \cos(r_m)]^\top \in \mathbb{R}^2$.
The token matrix is then constructed as
$\vT \triangleq [\vt_1,\ldots,\vt_M]^\top \in \mathbb{R}^{M\times 2}$. This transformation mitigates the phase-wrapping discontinuity and yields a continuous representation. Next, we employ a learnable linear projection matrix $\vE_{\mathrm{T}}\in\mathbb{R}^{2\times D_{\mathrm{emb}}}$ and obtain $\vT_{\mathrm{emb}} = \vT\vE_{\mathrm{T}}\in\mathbb{R}^{M\times D_{\mathrm{emb}}}$. This projection produces a dense $D_{\mathrm{emb}}$-dimensional representation for each token, which increases the model's expressive capacity and provides a suitable input dimension for subsequent self-attention and feed-forward layers. It is common practice to prepend a class token to $\vT_{\mathrm{emb}}$ that serves as a global sequence representation via its final hidden state~\cite{bert,vit}. In our experiments, however, incorporating a class token did not yield measurable performance gains; hence, we omit it in the proposed model. Instead, we utilize a \gls{gap} layer in the Output Head.} 

{We use learnable \gls{ap}-specific embeddings to distinguish \gls{ap} tokens, since Transformer encoders without positional information are permutation equivariant to token ordering~\cite{bishop_deepLearning}. Let $\vE_{\mathrm{AP}}\in\mathbb{R}^{M\times D_{\mathrm{emb}}}$ denote a learnable embedding matrix. The $m$-th token is assigned the $m$-th row of $\vE_{\mathrm{AP}}$, so that each \gls{ap} is associated with its own dedicated embedding vector. The entries of $\vE_{\mathrm{AP}}$ are trained jointly with the rest of the \gls{popt} network, analogously to the learned positional embeddings in~\cite{bert,vit}. 
Finally, the transformer input is formed by combining the projected phase tokens with the \gls{ap} embeddings, i.e., $\vH^{(0)}=\vT_{\mathrm{emb}}+\vE_{\mathrm{AP}}\in\mathbb{R}^{M\times D_{\mathrm{emb}}}$, which serves as the input to the Transformer Encoder.}

\subsubsection{Transformer Encoder}\label{subsubsec:encoder}
The encoder consists of $N_{\mathrm{enc}}$ stacked Transformer encoder blocks, each comprising \gls{ln}~\cite{ba2016layernormalization}, \gls{mha}, residual connections, and a token-wise \gls{mlp} with \gls{gelu} activation~\cite{gelu_ref}. 
{Since these transformer encoder operations are standard and well known~\cite{bishop_deepLearning}, we omit their detailed mathematical expressions for presentation brevity. Each encoder block has $N_{\mathrm{head}}$ attention heads with head dimension $d_h=D_{\mathrm{emb}}/N_{\mathrm{head}}$. The \gls{mha} module uses query, key, and value projections, followed by concatenation and an output projection back to $D_{\mathrm{emb}}$. The
token-wise MLP consists of two fully connected layers with hidden dimension $D_{\mathrm{ff}}$, i.e., $D_{\mathrm{emb}}\rightarrow D_{\mathrm{ff}} \rightarrow D_{\mathrm{emb}}$.} After $N_{\mathrm{enc}}$ layers, the transformer encoder output is $\vH^{(N_{\mathrm{enc}})}\in\mathbb{R}^{M\times D_{\mathrm{emb}}}$, which is subsequently provided to the Output Head.

\subsubsection{Output Head}\label{subsubsection_mlp_head}
{The encoder output $\vH^{(N_{\mathrm{enc}})}$ is first normalized using a \gls{ln} layer and then aggregated across the $M$ tokens using a \gls{gap} layer to obtain a single feature vector
\begin{equation}
    \vh_{\mathrm{gap}} = \mathrm{GAP}\!\left(\mathrm{LN}\!\left(\vH^{(N_{\mathrm{enc}})}\right)\right)\in\mathbb{R}^{D_{\mathrm{emb}}}.
\end{equation}
Here, \gls{gap} averages the $M$ token embeddings over the token dimension to produce a single global feature vector $\vh_{\mathrm{gap}}$. The coarse \gls{ue} position estimate is then obtained via a two-layer \gls{mlp} head with a hidden dimension of $D_{\mathrm{h}}$ and \gls{gelu} activation, expressed as
\begin{equation}
    \hat{\vx} = \vW_{2}\,\mathrm{GELU}\!\left(\vW_{1}\,\mathrm{LN}\!\left(\vh_{\mathrm{gap}}\right) + \vb_{1}\right) + \vb_{2}\in\mathbb{R}^{3},
\end{equation}
where $\vW_{1}\in\mathbb{R}^{D_{\mathrm{h}}\times D_{\mathrm{emb}}}$, $\vW_{2}\in\mathbb{R}^{3\times D_{\mathrm{h}}}$ are learnable weight matrices and $\vb_{1}\in\mathbb{R}^{D_{\mathrm{h}}}$, $\vb_{2}\in\mathbb{R}^{3}$ are learnable bias vectors.}

\vspace{-3mm}
\subsection{Benchmarks for Positioning}\label{subsection:benchmark_mlp}
To evaluate the performance of the proposed \gls{popt}, we consider two benchmark methods. As the first benchmark, we adopt the \gls{mlp}-based method in~\cite{11275391}, which directly maps the phase observations to the \gls{ue} position. To accommodate the 3D scenario considered in this work, the output layer dimension is extended to three, compared to the original 2D work in~\cite{11275391}. 

\textls[-3]{As a second benchmark, we consider the \gls{hi} method in~\cite{11274977}. The \gls{hi} method selects one \gls{ap} as a reference and forms differential phase measurements with the remaining \glspl{ap}, yielding differential distance constraints parameterized by the differential integer ambiguities. An \gls{mlp}-based estimator predicts these ambiguities, after which a \gls{gd} solver estimates the \gls{ue} position from the resulting constraints. Although it is referred to as hyperbola intersection, in the considered 3D setting the corresponding differential distance constraints are hyperboloids~\cite{s23146254}. Despite the fact that the phase superposition in our two-ray model \eqref{eq:received_signal} violates the single-path geometric assumptions underlying this method, it is included as an additional benchmark for completeness. Moreover, our recent method in~\cite{fatih_icassp} is excluded as a benchmark since it is inherently tailored to 2D positioning, where the intersection of two hyperbolas yields a unique solution. In contrast, in 3D the differential phase measurements define hyperboloids, whose pairwise intersections are generally not unique. Consequently, a direct extension to 3D would require at least three independent hyperboloid constraints and is therefore beyond the scope of this work. Finally, the methods proposed in~\cite{dey2025polo} are developed for 2D \gls{los}-only propagation scenarios and would require substantial modifications to be applicable to the scenario considered in this work. Therefore, they are not included as a benchmark.}

\vspace{-2mm}
\section{Complexity Analysis}\label{sec:complexity_analysis}
\textls[-4]{In this section, we analyze the inference complexity of the proposed \gls{popt} in Section~\ref{subsection:popt}, the \gls{egs} in Section~\ref{subsection:mle}, and the two benchmarks in Section~\ref{subsection:benchmark_mlp}, referred to as Benchmark \gls{mlp} and Benchmark \gls{hi}. The complexity is reported in terms of big-O ($\mathcal{O}$) scaling and the \gls{flop} count, a widely adopted metric~\cite{flop_motivation2_new, flop_count_reference, flop_count_reference_2, 11275391}. 
Specifically, we use the big-O notation to assess scalability with the number of \glspl{ap}, while the \gls{flop} count is used to quantify the computational burden relevant to real-time deployment. In our analysis, each elementary arithmetic operation is counted as one \gls{flop}, following the approach established in~\cite{flop_count_reference, flop_count_reference_2, flop_motivation2_new, 11275391}. We quantify the cost of the dominant matrix multiplications using the fact that multiplying an $(a_1\times a_2)$ matrix by an $(a_2\times a_3)$ matrix requires $a_1 a_3 (2a_2-1)\approx 2a_1 a_2 a_3$ \glspl{flop}, and scales as $\mathcal{O}(a_1 a_2 a_3)$. The computational cost of the fine estimation stage is neglected, since it is a common refinement step applied after the outputs of the considered coarse estimation methods and therefore does not affect their relative complexity comparison.}

\vspace{-3mm}
\subsection{Proposed POPT}
{The inference complexity of the \gls{popt} is dominated by the $N_{\mathrm{enc}}$ stacked Transformer encoder blocks; therefore, the costs of the Preprocessing stage and the Output Head are neglected.}
 In the \gls{mha} module, the per-head projections involve three matrix multiplications of size $(M \times D_{\mathrm{emb}})$ and $(D_{\mathrm{emb}} \times d_h)$. Each has complexity $\mathcal{O}(M D_{\mathrm{emb}} d_h)$ and costs approximately $2 M D_{\mathrm{emb}} d_h$ \glspl{flop}. Summed over three projections and $N_{\mathrm{head}}$ heads with $d_h=D_{\mathrm{emb}}/N_{\mathrm{head}}$, this yields $\mathcal{O}(M D_{\mathrm{emb}}^2)$ and $6 M D_{\mathrm{emb}}^2$ \glspl{flop}. The attention operation scales as $\mathcal{O}(M^2 d_h)$ per head and costs approximately $4 M^2 d_h$ \glspl{flop}; aggregated over all heads, it contributes $\mathcal{O}(M^2 D_{\mathrm{emb}})$ and $4 M^2 D_{\mathrm{emb}}$ \glspl{flop}. The output projection has complexity $\mathcal{O}(M D_{\mathrm{emb}}^2)$ and costs $2 M D_{\mathrm{emb}}^2$ \glspl{flop}. Consequently, the \gls{mha} module per layer scales as $\mathcal{O}(M^2 D_{\mathrm{emb}} + M D_{\mathrm{emb}}^2)$ with approximately $4 M^2 D_{\mathrm{emb}} + 8 M D_{\mathrm{emb}}^2$ \glspl{flop}.

{The token-wise \gls{mlp} block applies two fully connected layers to each of the $M$ tokens, yielding complexity $\mathcal{O}(M D_{\mathrm{emb}} D_{\mathrm{ff}})$ and approximately $4 M D_{\mathrm{emb}} D_{\mathrm{ff}}$ \glspl{flop}. The overhead of \gls{ln}, softmax, and \gls{gelu} is neglected as it is dominated by the matrix operations.}

By aggregating these components over $N_{\mathrm{enc}}$ layers, the total inference complexity of the \gls{popt} encoder is
\begin{equation}\label{eq:popt_bigO}
\mathcal{O}_{\text{POPT}} \!=\! \mathcal{O}\left( N_{\mathrm{enc}}\! \left( M^2 D_{\mathrm{emb}}\! +\! M D_{\mathrm{emb}} D_{\mathrm{ff}} \!+\! M D_{\mathrm{emb}}^2 \right) \right),
\end{equation}
and the total number of \glspl{flop} is approximately
\begin{equation}\label{eq:popt_flop}
\mathcal{F}_{\text{POPT}} \approx 2 N_{\mathrm{enc}} D_{\mathrm{emb}} \left( 2 M^2 + 4 M D_{\mathrm{emb}} + 2 M D_{\mathrm{ff}} \right).
\end{equation}
Although \eqref{eq:popt_bigO} and \eqref{eq:popt_flop} contain a quadratic term in $M$, in the considered setup and in many practical cases where $D_{\mathrm{emb}}$ and $D_{\mathrm{ff}}$ are typically set to several hundreds or thousands to maximize model capacity and much larger than $M$~\cite{11313538, 9022520}, the dominant contribution comes from the terms $M D_{\mathrm{emb}}^2$ and $M D_{\mathrm{emb}} D_{\mathrm{ff}}$. Therefore, in these practical settings, the terms linear in $M$ constitute the primary processing overhead, which supports the scalability of \gls{popt} and its suitability for real-time deployment in large-scale networks.

\vspace{-2mm}
\subsection{EGS/MLE Complexity}
The phase-only \gls{egs} in Section~\ref{subsection:mle} evaluates the cost function in \eqref{eq:egs} over a predefined search set $\mathcal{X}$. The number of grid points scales according to $\lvert \mathcal{X} \rvert = V / \delta_x^3$ for search space $V$ and grid resolution $\delta_x$ per side~\cite{dey2025polo}. Moreover, computing the \gls{ls} cost scales linearly with the number of \glspl{ap} for every $\vx \in \cX$ and the overall computational complexity is $\mathcal{O}_{\mathrm{EGS}}=\mathcal{O}\!\left(M V / \delta_x^3 \right)$. Computing \eqref{eq:egs} in every grid point requires predicting the phase, estimating the \gls{cpo} and computing the cost, requiring on average $B$ \glspl{flop} per \gls{ap}. For the considered problem, $B \approx 57$ \glspl{flop}, and the total number of \glspl{flop} is approximately $\mathcal{F}_{\mathrm{EGS}} \approx B M V / \delta_x^3$.

\vspace{-2mm}
\subsection{Complexity of Benchmark Methods}
For Benchmark \gls{mlp}, the inference complexity scales as 
$\mathcal{O}\!\left((M-1)A + A^2\right)$, where $A=128$, resulting in approximately $1.38\times 10^6$ \glspl{flop}~\cite{11275391}. For Benchmark \gls{hi}, the total inference cost is approximately 
$\mathcal{F}_{\mathrm{HI}} \approx 7.5\times 10^4 M + 256Q + 1.92\times 10^5$, 
where $Q$ denotes the number of output classes of the ambiguity estimator~\cite{11274977}. Since $Q$ grows linearly with $M$, Benchmark \gls{hi} has an overall inference complexity of $\mathcal{O}(M)$.

\vspace{-2mm}
\section{Numerical Results}\label{sec:num_results}
\textls[-4]{We consider a 3D evaluation area of size $10\,\mathrm{m}\times10\,\mathrm{m}\times4\,\mathrm{m}$ along the $x$-, $y$-, and $z$-axes, respectively. For the \gls{dmimo} network, $M=17$ \glspl{ap} are randomly deployed on the ceiling while ensuring a minimum separation distance between them to avoid unrealistic clustering. Following the 5G-NR \gls{srs} specifications in~\cite{3gpp_TS_38211}, the uplink pilot waveform is generated with a carrier frequency of $800$ MHz and a subcarrier spacing of $15$ kHz. The pilot allocation consists of $4$ resource blocks with comb factor $4$, corresponding to an effective pilot bandwidth of $W=180$ kHz. In addition, a single \gls{srs} reference symbol configuration is adopted following~\cite{3gpp_TS_38211}. The channel generation follows the two-ray model in Section~\ref{sec:system_model}, cf. \eqref{eq:received_signal}. The theoretical model assumes that the polarization type and ground material properties are perfectly known. Unless otherwise stated, the ground is assumed to be concrete. For the carrier frequency of $800$\,MHz, following~\cite{itu_reflection_ref}, the corresponding nominal complex relative permittivity is $\varepsilon_r \approx 5.24-\jmath0.0096$. The analytical model used for the \gls{peb} and the \gls{mle} assumes that the polarization type and the nominal ground material properties are known. In Section~\ref{sec:localization_results}, this assumption is intentionally relaxed by generating the received phase measurements with the nominal complex relative permittivity while using perturbed complex relative permittivity values in the \gls{mle} model. The considered transmit power levels, per the narrowband pilot structure described above, are $P \in \{-60,-50,-40,-30,-20,-10\}$\,dBm. The noise power is given by $\sigma_w^2 = k_B T_{\mathrm{tn}} W$, where $k_B = 1.38\times10^{-23}$ J/K denotes the Boltzmann constant and $T_{\mathrm{tn}} = 290$ K is the noise temperature. The receiver noise figure is set to $3$\,dB.}

In addition to the proposed \gls{popt}, we consider the \gls{mle} derived in Section~\ref{subsection:mle}, which relies on the theoretical model and is implemented via \gls{egs} over uniformly spaced candidate \gls{ue} locations in the evaluation area. Unless otherwise stated, the localization experiments use $1280$ \gls{mcs} realizations with random \gls{ue} locations and random \glspl{cpo}, a fixed transmit power of $P=-10\,\mathrm{dBm}$, $L=50$ \gls{ue} locations for \gls{gp} training, and an \gls{egs} grid spacing of $\delta_x=\lambda/8$. To avoid data leakage, the \gls{ue} locations used to acquire the limited \gls{gp} measurement dataset, the synthetic \gls{ue} locations used for \gls{popt} training and validation, and the \gls{ue} locations used for testing are sampled independently. 
For the proposed \gls{popt}, the \gls{mle}, and the two benchmarks, the coarse \gls{ue} position estimate is further refined using \gls{gd}, and the resulting output is referred to as the fine \gls{ue} position estimate.

\vspace{-4mm}
\subsection{Training of POPT}
As illustrated by Steps 3 and 4 in Fig.~\ref{fig:overall_framework}, the trained \gls{gp} model is first used offline to generate a synthetic labeled dataset $\tilde{\mathcal{D}} = \{(\tilde{\vx}_{n'}, \tilde{\vr}_{n'}) \mid n' = 1,\ldots, N_\textrm{AI} \}$ as presented in Section~\ref{sec:gp_regression}, which is then used for supervised training of \gls{popt}. The data is generated using $N_\textrm{AI}=1.125\times10^6$ random \gls{ue} positions which are drawn uniformly from the considered deployment area.
The proposed \gls{popt} uses an embedding dimension of $D_{\mathrm{emb}}=256$, $N_{\mathrm{enc}}=3$ stacked transformer encoder layers, $N_{\mathrm{head}}=4$ attention heads with per-head dimension $d_h=D_{\mathrm{emb}}/N_{\mathrm{head}}=64$, feed-forward dimension $D_{\mathrm{ff}}=1024$, and an Output Head hidden dimension of $D_{\mathrm{h}}=128$. Of the $N_{\mathrm{AI}}$ samples, $N_{\mathrm{tr}}=1\times10^6$ and $N_{\mathrm{val}}=0.125\times10^6$ are used for training and validation, respectively.
The network is trained by minimizing the \gls{mae} loss,
$\mathcal{L}_{\mathrm{MAE}}=\frac{1}{N_{\mathrm{tr}}}\sum_{n'=1}^{N_{\mathrm{tr}}}\left\lVert \tilde{\vx}_{n'}-\hat{\vx}_{n'}\right\rVert_1,$
where $\tilde{\vx}_{n'}$ and $\hat{\vx}_{n'}$ denote the ground-truth location and its estimate for sample $n'$, respectively. Optimization is performed using AdamW~\cite{adamw_ref} with a mini-batch size of $500$ for $N_{\mathrm{ep}}=120$ epochs, where a dropout rate of $0.01$ is applied to the input embeddings and encoder blocks. 
The \gls{lr} $\kappa$ follows a one-cycle schedule~\cite{onecycle_ref} with maximum value $\kappa_{\max}=10^{-3}$.
For each transmit power level and number of \gls{ue} locations $L$ used in \gls{gp} training, a separate \gls{popt} model is trained using the dataset $\tilde{\mathcal{D}}$ generated by the corresponding trained \gls{gp} model.

\vspace{-3mm} 
\subsection{Evaluation Metrics}\label{sec:evaluation_metrics}
The following metrics are computed for assessing the performance: the localization \gls{rmse}, modeling \gls{rmse}, and \gls{nlpd}. The localization accuracy is assessed using $\textrm{RMSE} = \sqrt{ \frac{1}{N_T} \sum_{n=1}^{N_T} \lVert \vx_{n} - \hat{\vx}_{n} \rVert_2^2}$, where $N_T$ is the number of test samples, $\vx_{n}$ and $\hat{\vx}_{n}$ denote the ground-truth location and its estimate for test location $n$, respectively.
The modeling accuracy is evaluated using $ \mathrm{RMSE} = \sqrt{\frac{1}{N_T M} \sum_{n=1}^{N_T} \sum_{m=1}^{M} e_{m,n}^2},$
where
$ e_{m,n} \triangleq \textrm{wrap}_{[-\pi,\pi)} \left(r_{m,n}-\mu_{m,n}\right) $
is the wrapped phase residual, $\mu_{m,n}$ is the predictive mean, and $r_{m,n}$ is the measured carrier phase for the $n$-th \gls{ue} location and $m$-th \gls{ap}. Lastly, the NLPD measures the modeling accuracy and is defined as
\begin{equation}
\mathrm{NLPD} = \frac{1}{2N_T M} \sum_{n=1}^{N_T} \sum_{m=1}^{M} \left( \frac{e_{m,n}^2}{\sigma_{m,n}^2} + \log(2\pi\sigma_{m,n}^2) \right),
\end{equation}
where $\sigma_{m,n}^2$ is the predictive variance.

\vspace{-3mm} 
\subsection{Modeling Results}
\begin{figure*}[!t]
  \vspace{-6mm}
  \centering
  \subfloat[True and estimated phase perturbations]{%
    \includegraphics[trim={0.0cm 0.1cm 0.9cm 0.4cm},clip, width=0.32\textwidth]{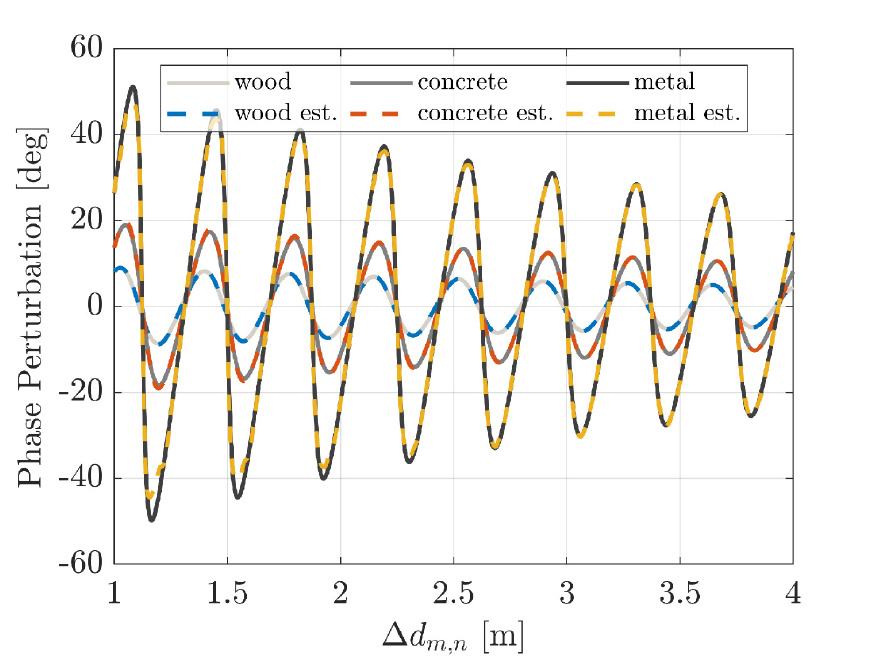}%
    \label{fig:example_signals}%
  }\hfill
  \subfloat[\gls{rmse} vs. $P$]{%
    \includegraphics[trim={0.0cm 0.1cm 0.9cm 0.4cm},clip, width=0.32\textwidth]{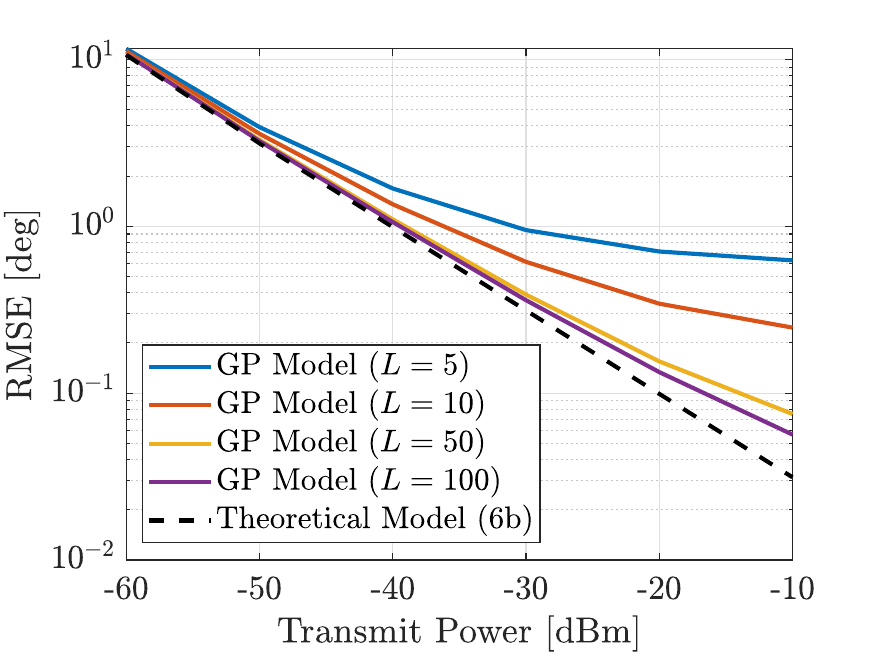}%
    \label{fig:RMSE_vs_PdBm_modeling}%
  }\hfill
  \subfloat[\gls{nlpd} vs. $P$]{%
    \includegraphics[trim={0.0cm 0.1cm 0.9cm 0.4cm},clip, width=0.32\textwidth]{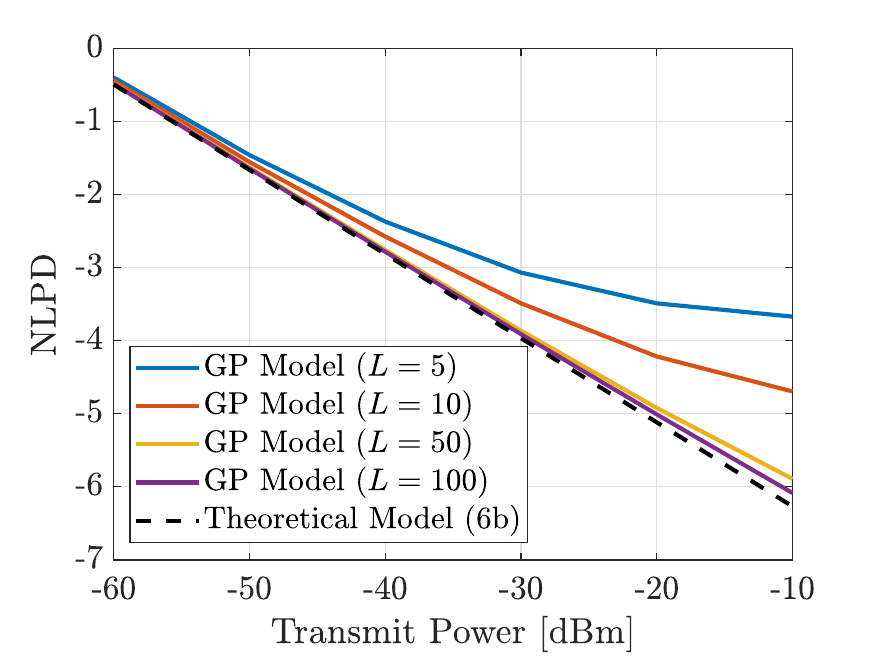}%
    \label{fig:NLPD_vs_PdBm_modeling}%
  }
  \vspace{1mm}
  \caption{Modeling results of the theoretical model in \eqref{eq:theta_eff_closedform} and proposed \gls{gp} model in \eqref{eq:gp_proposed_mean}-\eqref{eq:gp_proposed_variance}. In (a), the true and estimated phase perturbations (GP trained using $L=10$ UE locations) are shown for three different materials illustrating that the phase perturbations depend strongly on the reflecting material. In (b) and (c), \gls{rmse} and \gls{nlpd} illustrate how accurately the \gls{gp} models predict the mean and variance of the carrier phase measurements as a function of transmit power and the number of training samples when the ground material is concrete.}
  \label{fig:modeling_results}
  \vspace{-4mm}
\end{figure*}

Figure~\ref{fig:modeling_results} summarizes the modeling results and highlights three complementary aspects: (i) material-dependent phase perturbations, (ii) how well the \gls{gp} learns these perturbations, and (iii) how predictive uncertainty varies with transmit power and training data. Together, these results validate both the physical correctness and the predictive capabilities of the proposed \gls{gp} model.

\textls[-9]{Figure~\ref{fig:example_signals} illustrates that the \gls{nlos} phase-perturbation amplitude depends strongly on the reflecting material. First, metal produces the largest and most rapidly varying perturbations, as expected from its high-magnitude reflection coefficient. Therefore, the \gls{nlos} component strongly interferes with the \gls{los} signal, producing large quasi-periodic phase oscillations. Second, concrete and wood exhibit smaller variations, consistent with lower reflection coefficients and more lossy behavior. Finally, all materials exhibit quasi-periodic phase perturbations with a period equal to the wavelength.} 

Figures~\ref{fig:RMSE_vs_PdBm_modeling} and~\ref{fig:NLPD_vs_PdBm_modeling} evaluate how accurately the \gls{gp} models predict the mean and variance of the carrier phase measurements as a function of transmit power and the number of training samples. Here, the theoretical model refers to the analytical two-ray carrier phase model in \eqref{eq:carrier_phase_measurements}, where the effective phase $\theta_m(\vx)$ is obtained from \eqref{eq:theta_eff_closedform} through the reflection coefficient in \eqref{eq:gamma}, under the assumption that the relative permittivity $\varepsilon_r$ and polarization mode are known. In general, a higher transmit power yields lower values of \gls{rmse} and \gls{nlpd}, as expected, because the measurement noise decreases as \gls{snr} increases. In addition, the accuracy improves as the number of training samples grows. With $L\mathord{=}100$, the \gls{gp} model nearly matches the theoretical model, indicating that its mean and covariance functions are well designed and that the proposed approach can approximate carrier phase measurements with high fidelity. 
At high transmit powers, even with very few training samples ($L\mathord{=}5$), the \gls{gp} model yields an estimation error below one degree, showing that the joint formulation, in which all \glspl{ap} share the same model, is effective in practice.

\textls[-3]{Table~\ref{tab:material_comparison} compares the \gls{rmse} and \gls{nlpd} of the theoretical and \gls{gp} models across five common materials~\cite{itu_reflection_ref}, using $L\mathord{=}100$ \gls{ue} locations for training and $P\mathord{=}-10\,\text{dBm}$. For low-reflectivity materials (wood, plasterboard, concrete), the \gls{rmse} and \gls{nlpd} are nearly identical, indicating near-theoretical accuracy with the proposed \gls{gp} model. 
In contrast, for high-reflectivity materials (glass, metal), the \gls{gp} model's \gls{rmse} and \gls{nlpd} increase relative to the theoretical model.
This trend is physically intuitive: stronger reflections induce larger and sharper quasi-periodic phase fluctuations, which are harder to learn from limited data.}

Overall, Fig.~\ref{fig:modeling_results} shows that the proposed \gls{gp} model captures the quasi-periodic structure of the two-ray model, with perturbation magnitudes that depend on the reflecting material and incidence angle, thereby supporting its physical interpretability. Moreover, the proposed \gls{gp} model attains high accuracy with relatively few training samples, and the joint formulation improves performance as $L$ increases, approaching the theoretical model at $L=100$. These results support the proposed \gls{gp} model as a physics-informed machine-learning approach for phase-only positioning in \gls{dmimo} systems.

\begin{table}[t!]
\centering
\caption{\textsc{\gls{rmse} in degrees and \gls{nlpd} for different materials with $M=17$, $L=100$, and $P=-10\,\mathrm{dBm}$. The material-dependent relative permittivities at $800$ MHz are taken from~\cite{itu_reflection_ref}.}}

\label{tab:material_comparison}
\renewcommand{\arraystretch}{1.0}
\begin{tabular}{*{5}{c}}
\hline
& \multicolumn{2}{c}{\textbf{Theoretical}} & \multicolumn{2}{c}{\textbf{\gls{gp} Model}} \\  
\cline{2-5}
\textbf{Material} & \gls{rmse} & \gls{nlpd} & \gls{rmse} & \gls{nlpd}  \\
\hline
wood & $0.03$ & $-6.27$ & $0.04$ & $-6.16$ \\ 
plasterboard & $0.03$ & $-6.27$ & $0.05$ & $-6.14$ \\ 
concrete & $0.03$ & $-6.27$ & $0.06$ & $-6.08$ \\ 
glass & $0.04$ & $-6.26$ & $0.14$ & $-5.69$ \\ 
metal & $0.05$ & $-6.26$ & $0.29$ & $-5.28$ \\ 
\hline
\end{tabular}
\vspace{3mm}
\end{table}

\vspace{-3mm}
\subsection{Localization Results}\label{sec:localization_results}
\begin{figure*}[!t]
  \vspace{-3mm}
  \centering
  \subfloat[Position Errors vs. Transmit Power]{%
    \includegraphics[trim={0.0cm 0.0cm 0.9cm 0.4cm},clip, width=0.32\textwidth]{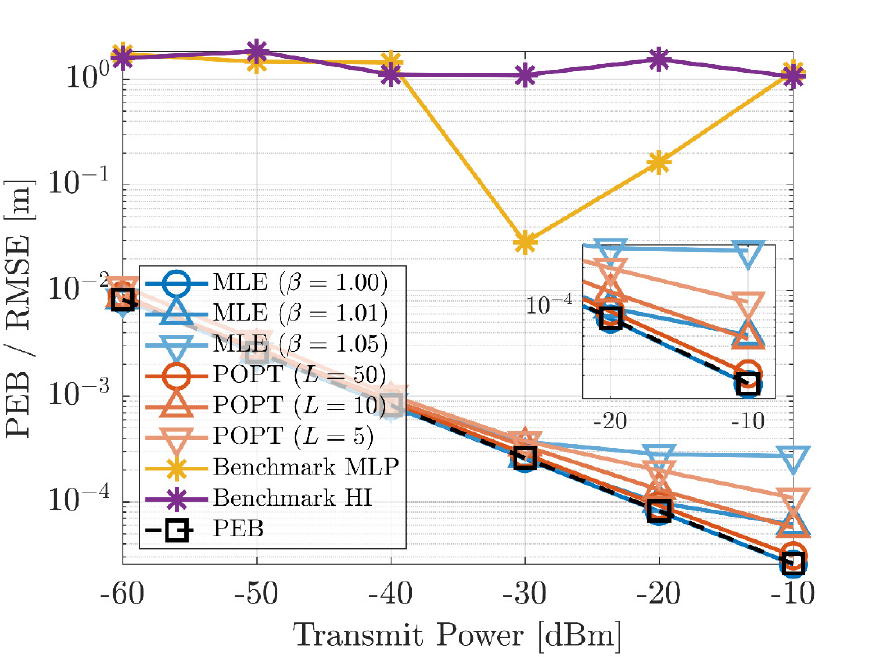}%
    \label{fig:peb_and_rmse_vs_transmit_power}%
  }\hfill
  \subfloat[Empirical CDFs for $P\mathord{=}-10\,\text{dBm}$]{%
    \includegraphics[trim={0.0cm 0.0cm 0.9cm 0.4cm},clip, width=0.32\textwidth]{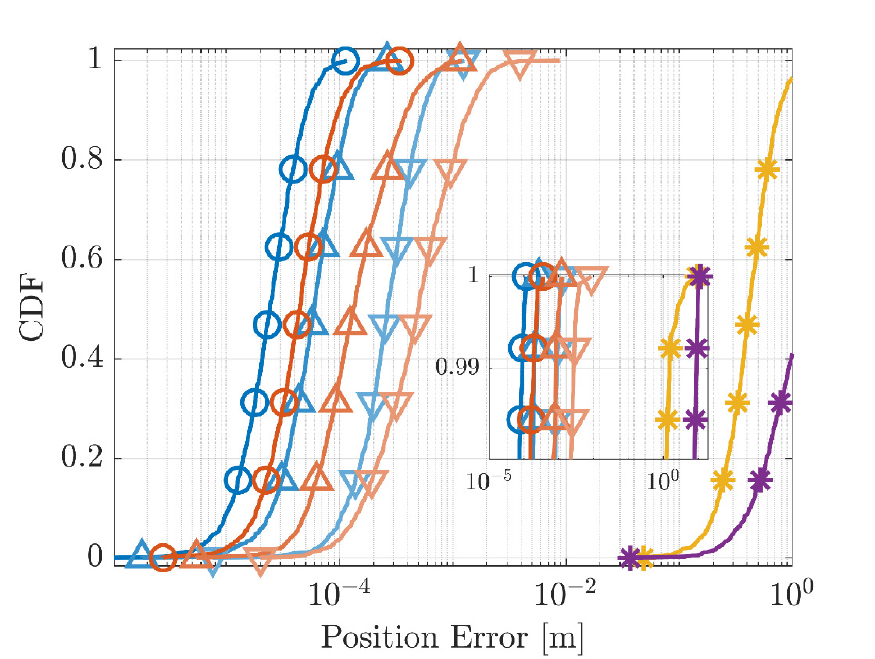}%
    \label{fig:position_error_cdfs}%
  }\hfill
  \subfloat[Coarse and Fine Position Errors]{%
    \includegraphics[trim={0.0cm 0.0cm 0.9cm 0.4cm},clip, width=0.32\textwidth]{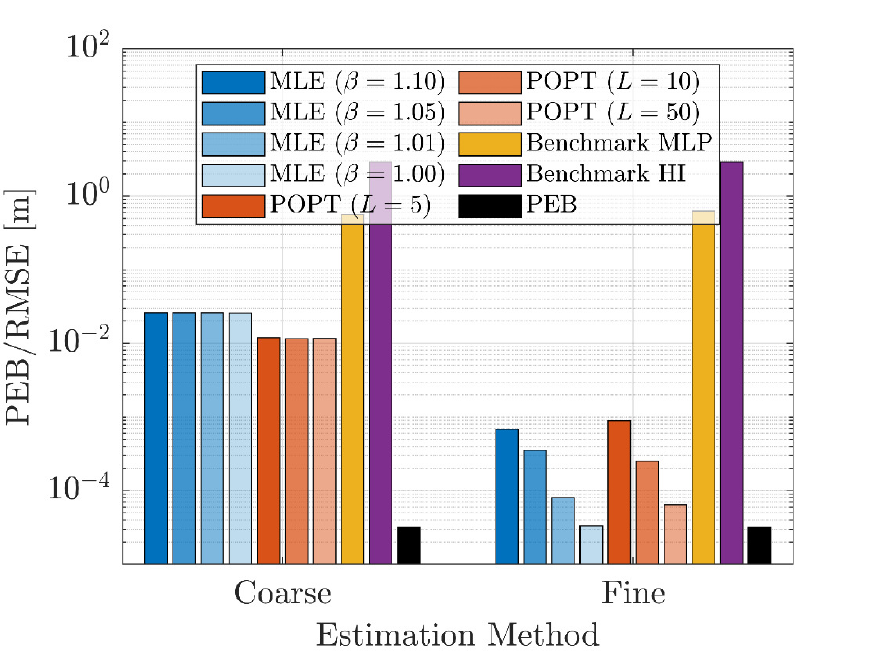}%
    \label{fig:theoretical_vs_gp_bar_plot}%
  }
  \vspace{1mm}
  \caption{Positioning performance of the proposed and benchmark methods under different transmit powers, relative permittivity mismatch levels, and \gls{gp} training data sizes. (a) \gls{peb} and \gls{rmse} of the fine \gls{ue} position estimates versus transmit power for fixed \gls{ue} location and \gls{cpo}. For the \gls{mle}, the \gls{egs} grid spacing is fixed to $\delta_x=\lambda/8$, while the mismatch level is controlled by $\beta$. For \gls{popt}, one curve per $L$ is shown because the \gls{gp}-based model adapts to the corresponding propagation condition. (b) Empirical \glspl{cdf} of the fine estimation error for $1280$ random \gls{ue} locations and random \glspl{cpo} at $P\mathord{=}-10\,\text{dBm}$ (legend of the plot is the same as in (a)). (c) Mean \gls{peb} and \gls{rmse} for coarse and fine estimates, with $P\mathord{=}-10\,\text{dBm}$, illustrating the sensitivity of \gls{mle} to relative permittivity mismatch and the impact of the number of \gls{gp} training locations for the proposed approach.}
  \label{fig:positioning_results}
  \vspace{-3mm}
\end{figure*}

\textls[-3]{Figure~\ref{fig:positioning_results} evaluates the localization accuracy of the proposed and benchmark methods under relative permittivity mismatch. This mismatch is motivated by the fact that even closely related reflecting materials can exhibit noticeable differences in their electrical properties. For example, the ITU-R material-parameter model in~\cite{itu_reflection_ref} yields approximately $\varepsilon_r=5.24-\jmath0.0096$ for concrete and $\varepsilon_r=4.83-\jmath0.18$ for asphalt concrete at $800$ MHz, corresponding to around an $8\%$ difference in the magnitude of the complex relative permittivity. Motivated by this, the carrier phase measurements are generated and the corresponding \gls{peb} is computed using a fixed complex relative permittivity $\varepsilon_r$, while the analytical \gls{mle} assumes a perturbed value $\varepsilon_r^{(\beta)} = \beta \varepsilon_r, \textrm{with } \beta \in \{1.00, 1.01, 1.05, 1.10\}$. Therefore, $\beta=1.00$ corresponds to the matched case, while $\beta>1.00$ represents increasing mismatch in the material parameter assumed by the \gls{mle}. The different \gls{mle} curves in Fig.~\ref{fig:positioning_results} therefore quantify the sensitivity of the model-based estimator to imperfect knowledge of the relative permittivity. Since the carrier phase measurements are generated using the same fixed relative permittivity for all cases, no separate \gls{popt} curves are shown for different $\beta$ values; the \gls{popt} curves differ only in the GP training size $L$.}

Figure~\ref{fig:peb_and_rmse_vs_transmit_power} shows the \gls{peb} and the \gls{rmse} of the fine \gls{ue} position estimates for different transmit power levels. The shown \gls{peb} corresponds to the matched material model, i.e., $\beta=1.00$. The shown results are obtained by averaging over $1280$ noise realizations and keeping the \gls{ue} location and \gls{cpo} fixed to $\vx = [6.5,\, 5.0, \, 2.0]^\top \,\text{m}$ and $\phi=\pi/2\, \text{rad}$, respectively. The \gls{peb} decreases monotonically with increasing transmit power, reflecting the reduced phase noise at higher \gls{snr} levels. When the material model is matched, i.e., $\beta=1.00$, the \gls{mle} closely follows the \gls{peb}, confirming that the adopted \gls{egs} followed by \gls{gd} refinement can attain the theoretical accuracy under perfect model knowledge. However, even a small mismatch in $\varepsilon_r$ degrades the \gls{mle} performance. This degradation becomes particularly visible at high transmit powers, where the random phase noise is small and the modeling error induced by the incorrect reflection coefficient becomes the dominant error source. Conversely, the proposed \gls{popt} approach maintains high accuracy because the \gls{gp} model does not require explicit knowledge of $\varepsilon_r$; instead, it learns the phase perturbation from the measurement dataset. As $L$ increases, the \gls{gp} model becomes more accurate and the corresponding \gls{popt} localization performance improves, with $L\!=\!50$ approaching the \gls{peb} over the considered transmit power range. {On the other hand, Benchmark \gls{mlp} and Benchmark \gls{hi} exhibit degraded localization performance, as they are originally designed for 2D \gls{los}-only propagation and therefore do not explicitly account for the 3D two-ray multipath effects.}

Figure~\ref{fig:position_error_cdfs} illustrates the empirical positioning error \glspl{cdf} of the fine estimates under the default localization setting. The \glspl{cdf} show that the matched \gls{mle} case with $\beta=1.00$ provides highly concentrated errors, while increasing $\beta$ shifts the \gls{mle} distributions to the right and produces larger tail errors. The \glspl{cdf} demonstrate that increasing the number of \gls{gp} training locations, $L$, improves both the average accuracy and the empirical tail behavior of the proposed method. With $L\mathord{=}50$, $99\,\%$ of the positioning errors remain below $0.191\,\mathrm{mm}$, and the maximum observed error is $0.352\,\mathrm{mm}$. This tail behavior is particularly critical for safety-critical applications, where worst-case performance guarantees are essential. {Consistent with Fig.~\ref{fig:peb_and_rmse_vs_transmit_power}, the benchmark methods exhibit substantially larger errors, highlighting their limited suitability for the considered 3D two-ray propagation scenario.}

Finally, Fig.~\ref{fig:theoretical_vs_gp_bar_plot} reports the mean \gls{peb} and \gls{rmse} values for both the coarse and fine \gls{ue} position estimates under the default localization setting. From the coarse estimation results, it can be observed that the \gls{egs} accuracy remains almost unaffected by the different $\beta$ values. After \gls{gd} refinement, however, the effect of the relative permittivity mismatch becomes clear: the matched \gls{mle} case approaches the \gls{peb}, whereas larger $\beta$ values lead to a noticeable increase in the fine positioning error. This shows that the \gls{gd} refinement is highly sensitive to the correctness of the analytical reflection model. For the proposed approach, the coarse estimates are already accurate for all considered $L$ values, and the fine estimate improves as more \gls{gp} training locations are used. Overall, Fig.~\ref{fig:positioning_results} demonstrates the key advantage of the proposed physics-informed learning pipeline: the \gls{gp} model can adapt to environment-dependent phase perturbations from a small number of measurements, whereas the analytical \gls{mle} is vulnerable to even modest mismatches in the complex relative permittivity.

\vspace{-3mm}
\subsection{Computational Complexity Results}
Based on the complexity analysis in Section~\ref{sec:complexity_analysis}, Table~\ref{tab:complexity_runtime} summarizes the \gls{flop} counts, measured inference times, and fine estimate \gls{rmse} values for all considered methods. The reported runtimes correspond to the coarse inference stages analyzed in Section~\ref{sec:complexity_analysis} and do not include the \gls{gd} refinement stage, since this refinement is applied to all considered methods and is therefore excluded to enable a fair comparison of the method-specific inference costs. 
The runtimes of the proposed \gls{popt}, Benchmark \gls{mlp}, and Benchmark \gls{hi} are measured in Python 3.13.7 with PyTorch 2.12.1, while those of the \gls{mle} are measured in MATLAB R2024a, both on an Intel Core i7-13700H CPU.
The fine estimate \gls{rmse} values for the proposed method, the benchmarks, and the matched \gls{mle} with $\delta_x=\lambda/8$ are consistent with Fig.~\ref{fig:theoretical_vs_gp_bar_plot}. The additional \gls{mle} rows with $\delta_x=\lambda/2$ and $\delta_x=\lambda/4$ are included to illustrate the complexity--resolution trade-off of the \gls{egs} stage under the matched material model.

\textls[-4]{Table~\ref{tab:complexity_runtime} exposes a clear accuracy--complexity trade-off. \gls{mle} requires a grid spacing no coarser than $\delta_x\mathord{=}\lambda/8$ to approach the \gls{peb}, which costs $3771.04\times 10^6$ \glspl{flop} and a mean runtime of $3716.05$ ms. In contrast, the proposed \gls{popt}, trained using synthetic data generated by the \gls{gp} model learned from only $L\mathord{=}50$ \gls{ue} locations, attains comparable accuracy at a cost of only $81.10\times 10^6$ \glspl{flop} and $0.95$ ms runtime, corresponding to reductions of approximately $46.5$ times in \gls{flop} count and $3.9\times 10^{3}$ times in runtime, i.e., $1.7$ and $3.6$ \emph{orders of magnitude}, respectively. The resulting sub-millisecond {coarse} inference time of \gls{popt} is below the most stringent $[1\text{--}10]$ ms end-to-end service latency target reported for real-time digital twins in~\cite{3gpp_tr_22_870}. By comparison, \gls{mle} fails to meet this requirement even at the coarsest grid spacing $\delta_x\mathord{=}\lambda/2$, for which its mean runtime is already $64.03$ ms. Benchmark \gls{mlp} and Benchmark \gls{hi} achieve runtimes of $0.40$ ms and $20.33$ ms, respectively, but neither reaches the fine estimate accuracy of the proposed approach. Moreover, despite having a relatively low \gls{flop} count ($1.56\times 10^6$), Benchmark \gls{hi} exhibits a higher runtime than both Benchmark \gls{mlp} and \gls{popt} due to its internal iterative \gls{gd}-based refinement~\cite{11274977}.}

Table~\ref{tab:complexity_runtime} further reveals a discrepancy between the \gls{flop} count and the measured runtime when comparing \gls{popt} with \gls{mle} at $\delta_x\mathord{=}\lambda/8$: the \gls{flop} ratio is only about $46.5$, whereas the runtime ratio reaches about $3.9\times 10^{3}$. This gap is expected, since the \gls{flop} count alone does not capture hardware efficiency. The inference of \gls{popt} is dominated by dense matrix multiplications, which are highly optimized in modern numerical libraries and hardware, whereas \gls{mle} sweeps a large candidate set and therefore incurs substantial memory-access overhead and less efficient loop execution. In addition, the simplified \gls{flop} accounting of \gls{mle} treats each trigonometric and square-root evaluation as a single \gls{flop}, while in practice such operations may require several to tens of elementary operations per call.
\begin{table}[t]
\caption{Comparison of \gls{flop} counts, inference times, and fine \glspl{rmse}. Inference time is reported in milliseconds as mean $\pm$ standard deviation over $1280$ Monte Carlo realizations.}
\label{tab:complexity_runtime}
\centering
\renewcommand{\arraystretch}{1.05}
\setlength{\tabcolsep}{4pt}
\begin{tabular}{lccc}
\hline
\shortstack{\textbf{Estimation}\\\textbf{Method}} 
& \shortstack{\textbf{\glspl{flop}}\\\textbf{($\boldsymbol{\times 10^6}$)}} 
& \shortstack{\textbf{Inference time}\\\textbf{(ms)}} 
& \shortstack{\textbf{Fine RMSE}\\\textbf{[cm]}}\\
\hline
MLE ($\delta_x=\lambda/2$)  & $58.92$   & $64.03 \pm 2.26$      & $2.23\times 10^{2}$ \\
MLE ($\delta_x=\lambda/4$)  & $471.38$  & $480.14 \pm 7.18$   & $1.90\times 10^{1}$ \\
MLE ($\delta_x=\lambda/8$)  & $3771.04$ & $3716.05 \pm 26.85$ & $3.34\times 10^{-3}$ \\
Benchmark MLP              & $1.38$    & $0.40 \pm 0.17$      & $6.27\times 10^{1}$ \\
Benchmark HI               & $1.56$    & $20.33 \pm 0.61$      & $2.89\times 10^{2}$ \\
Proposed POPT                  & $81.10$   & $0.95 \pm 0.20$      & $6.50\times 10^{-3}$ \\
\hline
\end{tabular}

\vspace{1mm}
\footnotesize{For reference, the mean \gls{peb} is $3.14\times 10^{-3}$ cm, as shown in Fig.~\ref{fig:theoretical_vs_gp_bar_plot}.}
\vspace{3mm}
\end{table}

\vspace{-2mm}
\section{Conclusion}\label{sec:conclusion}
\textls[-3]{In this article, we developed and proposed a physics-informed deep learning framework for phase-only positioning in uplink phase-coherent \gls{dmimo} systems under dominant multipath scenarios. We proposed a \gls{gp}-based carrier phase model that captures the quasi-periodic phase perturbations induced by ground reflections and enables data-efficient generation of high-quality synthetic training samples. Building on this model, we introduced \gls{popt}, an encoder-only transformer, while also deriving the fundamental \gls{peb} and developing the MLE for the considered phase-only positioning problem. The numerical results show that the proposed \gls{gp} model attains high fidelity with relatively few training samples and that the localization accuracy improves consistently as the \gls{gp} training set grows. In particular, with 50 training locations, the proposed approach achieves near-\gls{peb} performance and outperforms the two considered benchmarks. Furthermore, the results show that the analytical \gls{mle} is vulnerable to even modest errors in the assumed relative permittivity, whereas the proposed \gls{gp}-based approach avoids explicitly relying on this parameter by learning the phase perturbation from measurement data.
Moreover, compared with the \gls{mle} relying on exhaustive grid search, the proposed \gls{popt} reduces the \gls{flop} complexity and inference time by approximately 1.7 and 3.6 orders of magnitude, respectively.} 
\textls[-3]{These results highlight the potential of the proposed framework for computationally feasible, low-latency and high-accuracy positioning in future phase-coherent distributed wireless networks. Future work will extend the proposed framework to more realistic propagation environments, including orientation-dependent antenna patterns, spatially heterogeneous reflecting surfaces with non-constant relative permittivity, and more general multipath propagation beyond the two-ray model. Furthermore, extending the proposed single-snapshot approach to dynamic scenarios with moving \glspl{ue} is an important direction, where multi-snapshot processing and sequential learning-based tracking methods can be leveraged.}

\vspace{-0mm}

\appendices
\section*{Appendix: Jacobian Matrix Derivation}\label{sec:jacobian_matrix_derivation}
In the following, we derive the Jacobian matrix, defined as $[\vH_\vx(\vtheta)]_{m} = \partial \theta_m / \partial \vx$. By utilizing the definition of complex logarithm $\log z = \ln{\vert z \vert} + \jmath \arg{z} \Leftrightarrow \arg{z} = \Im\left\{ \log z \right\}$, the $m$th row of the Jacobian matrix can be written as 
\begin{equation}\label{eq:phase_diff}
\!\!\! \frac{\partial \theta_m}{\partial \vx} \!=\! \Im \left\{\frac{\sum_{i=1}^2 \! \left( \frac{\partial \alpha_m^{(i)}}{\partial \vx} \!-\! \jmath \alpha_m^{(i)} \frac{\partial \theta_m^{(i)}}{\partial \vx} \right)\exp(-\jmath\theta_m^{(i)}) }{\sum_{i=1}^2 h_m^{(i)}(\vx)}\right\}.    
\end{equation}
The partial derivatives for the amplitude and phase of the \gls{los} path can be straightforwardly obtained as
\begin{equation}
\frac{\partial \alpha_m^{(1)}}{\partial \vx} = \frac{\lambda}{4\pi} \frac{\vdeltam^\top}{\Vert \vdeltam \Vert^3} \quad \text{and} \quad \frac{\partial \theta_m^{(1)}}{\partial \vx} = -\frac{2\pi}{\lambda} \frac{\vdeltam^\top}{\Vert \vdeltam \Vert},
\end{equation}
respectively, where $\vdeltam = \vp_m-\vx$. Furthermore, regarding the ground-reflected path, the corresponding partial derivatives of the amplitude and phase are given as
\begin{subequations}
\begin{align}
\frac{\partial \alpha_m^{(2)}}{\partial \vx} &= \frac{\lambda}{4\pi \Vert \vdeltamtilde \Vert} \left(\frac{\vert \Gamma_m(\vx)\vert \vdeltamtilde^\top}{\Vert \vdeltamtilde \Vert^2} + \frac{\partial \vert \Gamma_m(\vx)\vert}{\partial \vx}\right) \text{  and} \\
\frac{\partial \theta_m^{(2)}}{\partial \vx} &= -\frac{2\pi}{\lambda} \frac{\vdeltamtilde^\top}{\Vert \vdeltamtilde \Vert} - \frac{\partial \angle \Gamma_m(\vx)}{\partial \vx},
\end{align}
\end{subequations}
where $\vdeltamtilde = \tilde{\vp}_m-\vx$, and $\Gamma_m(\vx)$ is the reflection coefficient. Considering the complex logarithm definition, used in \eqref{eq:phase_diff}, while exploiting the identities $\vert z \vert = \sqrt{z z^*}$ and $z + z^* = 2\Re\left\{ z \right\}$, the partial derivatives for the amplitude and phase of the reflection coefficient can be formulated as
\begin{subequations}
\begin{align}
\frac{\partial \vert \Gamma_m(\vx) \vert}{\partial \vx} &= \Re \left\{\frac{\Gamma_m^*(\vx)}{\vert \Gamma_m(\vx) \vert} \frac{\partial \Gamma_m(\vx)}{\partial \vx} \right\}\text{ and} \\
\frac{\partial \angle \Gamma_m(\vx)}{\partial \vx} &= \Im \left\{\frac{1}{\Gamma_m(\vx)} \frac{\partial \Gamma_m(\vx)}{\partial \vx} \right\},
\end{align}
\end{subequations}
respectively. After some algebraic manipulations, the partial derivative of the reflection coefficient can be written as
\begin{align}
\frac{\partial \Gamma_m(\vx)}{\partial \vx} = \frac{2(1-\varepsilon_r)\sin(\gamma_m(\vx))}{\zeta(\vx)(\cos(\gamma_m(\vx)) + \zeta(\vx))^2} \frac{\partial \gamma_m(\vx)}{\partial \vx},
\end{align}
where 
$\zeta(\vx) = \sqrt{\varepsilon_r - \sin^2(\gamma_m(\vx))}$. Let $\mathbf{e}_3 = [0,0,1]^\top$; then the partial derivative of the incidence angle can be expressed as
\begin{equation}
\frac{\partial \gamma_m(\vx)}{\partial \vx}
= -\frac{(z_m+z_{i})\vdeltamtilde^\top + \left(d_m^{(2)}\right)^2 \mathbf{e}_3^\top}{\left(d_m^{(2)}\right)^3 \sqrt{1-\cos^2(\gamma_m(\vx))}}.
\end{equation}

\bibliographystyle{IEEEtran}
\bibliography{references}

\end{document}